\documentclass[journal]{IEEEtran}

\usepackage{cite}
\usepackage{graphicx}
\usepackage{url}
\usepackage{amsmath}
\usepackage{amssymb}
\usepackage[caption = false, font=footnotesize]{subfig}
\usepackage{float}
\usepackage{booktabs} 

\usepackage{array}
\usepackage{color}
\usepackage{tabularx}
\usepackage{longtable}
\usepackage{tabu}
\usepackage{multirow}
\usepackage{bm}
\usepackage{times}
\usepackage{threeparttable}

\newcolumntype{L}[1]{>{\raggedright\let\newline\\\arraybackslash\hspace{0pt}}m{#1}}
\newcolumntype{C}[1]{>{\centering\let\newline\\\arraybackslash\hspace{0pt}}m{#1}}
\newcolumntype{R}[1]{>{\raggedleft\let\newline\\\arraybackslash\hspace{0pt}}m{#1}}

\newcommand{\ie}{\textit{i}.\textit{e}.}
\newcommand{\eg}{\textit{e}.\textit{g}.}

\DeclareMathOperator*{\argmin}{argmin}

\definecolor{grey}{RGB}{130,130,130}
\definecolor{black}{RGB}{0,0,0}
\definecolor{red}{RGB}{255,0,0}

\ifCLASSINFOpdf

\else

\fi

\makeatletter
\def\ps@IEEEtitlepagestyle{%
  \def\@oddfoot{\mycopyrightnotice}%
  \def\@evenfoot{}%
}
\def\mycopyrightnotice{%
 \resizebox{\textwidth}{!}{{\footnotesize Copyright {\copyright} 20xx IEEE. Personal use of this material is permitted. However, permission to use this material for any other purposes must be obtained from the IEEE by sending an email to \underline{pubs-permissions@ieee.org}.}}
}
\makeatother

\begin{document}
%
\title{Blindly Assess Quality of In-the-Wild Videos via Quality-aware Pre-training and Motion Perception}
%
%
%

\author{Bowen~Li,
        Weixia~Zhang,~\IEEEmembership{Member,~IEEE,}
        Meng~Tian,~\IEEEmembership{Member,~IEEE,}
        Guangtao~Zhai,~\IEEEmembership{Senior Member,~IEEE,}
        and~Xianpei~Wang
\thanks{This work was supported in part by the National Natural Science Foundation of China under Grant 61901262, 52177109, and 51707135, in part by the Key R\&D Program of Hubei Province, China under Grant 2020BAB109, in part by Fundamental Research Funds for the Central Universities, China under Grant 2042019kf1014.}
\thanks{Bowen Li, Meng Tian, and Xianpei Wang are with the Electronic Information School, Wuhan University, Wuhan 430072, China (e-mail: bornlee@whu.edu.cn; mengtian@whu.edu.cn; xpwang@whu.edu.cn).}
\thanks{Weixia Zhang, and Guangtao Zhai are with the MoE Key Laboratory of Artificial Intelligence, AI Institute, Shanghai Jiao Tong University, Shanghai 200240, China (e-mail: zwx8981@sjtu.edu.cn; zhaiguangtao@sjtu.edu.cn).}
}

\markboth{}%
{Shell \MakeLowercase{\textit{et al.}}: Bare Demo of IEEEtran.cls for Journals}

\maketitle

\begin{abstract}
Perceptual quality assessment of the videos acquired in the wilds is of vital importance for quality assurance of video services. The inaccessibility of reference videos with pristine quality and the complexity of authentic distortions pose great challenges for this kind of blind video quality assessment (BVQA) task. Although model-based transfer learning is an effective and efficient paradigm for the BVQA task, it remains to be a challenge to explore~\textit{what} and~\textit{how} to bridge the domain shifts for better video representation. In this work, we propose to transfer knowledge from image quality assessment (IQA) databases with authentic distortions and large-scale action recognition with rich motion patterns. We rely on both groups of data to learn the feature extractor and use a mixed list-wise ranking loss function to train the entire model on the target VQA databases. Extensive experiments on six benchmarking databases demonstrate that our method performs very competitively under both individual database and mixed databases training settings. We also verify the rationality of each component of the proposed method and explore a simple ensemble trick for further improvement.
\end{abstract}


%
\IEEEpeerreviewmaketitle
\section{Introduction}\label{Sec:Introduction}
\IEEEPARstart{T}{he} Global Internet Phenomena declares that video streaming has already made up more than 60\% of the whole Internet traffic~\cite{GIP2019}. And a report by a Cisco project shows that online videos will account for more than 82\% of all consumer Internet traffic by 2022~\cite{cisco20172022forecast}. Confronted With various video providers, consumers always expect favorable quality-of-experience (QoE)~\cite{duanmu2016quality} when they are paying for these video services. Therefore, it is of high importance to develop reliable video quality assessment (VQA) models to ensure the quality of video services.

Because humans are the ultimate receivers of videos, the most promising VQA methodology is subjective quality testing. However, conducting such testing is labor-intensive and time-consuming, resulting in poor scalability to large-scale applications. As an alternative, objective VQA aims at automatically predicting the quality of videos. Objective VQA includes three categories: full-reference (FR) VQA, reduced-reference (RR) VQA, and no-reference/blind (NR/B) VQA. FR(RR)-VQA methods (partially) rely on non-distorted videos for making quality predictions, thus are not applicable for applications where the pristine videos are inaccessible or even not existing \cite{tu2021regression}. Consequently, increasing attention has been paid to BVQA over the past years.

Early BVQA methods were mainly developed for specific distortion types such as transmission and compression~\cite{amer2005fast, valenzise2011no, sogaard2015no}. Although a plethora of general-purpose BVQA models was developed subsequently, they were still designed for handling synthetic distortions using hand-crafted features~\cite{mittal2012no, saad2014blind, mittal2016completely}. These methods usually struggle for VQA in the wild~\cite{li2019quality}, where the distortions are naturally introduced during the video acquisition. Such authentic distortions may originate from various factors, including amateurish photographing, low-end camera devices, poor shooting environments, inappropriate post-processing, etc.

Due to the remarkable representation learning capability, deep neural networks (DNNs) have presented their promises in various vision applications over the past years. However, direct applications of the powerful DNNs for VQA tasks usually suffer from two main challenges: 1), prohibitively high computational complexity and memory consumption for processing the whole videos (usually with high spatial resolutions); 2), insufficient corpus with human-annotated quality labels for training effective DNNs from scratch. Recent work may suggest leveraging a large number of videos with pseudo-labels~\cite{zhang2019blind, wu2021no} to train 3D models from scratch. While methods of this kind handle videos with synthetic distortions (\eg, compression, transmission errors) well, they are found to present sub-optimal generalizability to in-the-wild videos due to the distributional shifts~\cite{spatio2021liu}. To mitigate the above issues, previous methods follow a paradigm to employ pre-trained DNNs on large-scale image classification databases~\cite{deng2009imagenet} to extract frame-level features~\cite{li2019quality, li2021unified, tu2021rapique}. The philosophy behind this paradigm is straightforward because videos are composed of sequences of images. Despite being empirically effective and efficient for VQA in the wild, this paradigm inevitably confronts the problem of distributional shifts~\cite{zhou2021domain} between the source domains (\eg, image classification) and the target domains (in-the-wild VQA), resulting in sub-optimal feature representation. In addition to the frame-level spatial features, motion information also plays an important role in human perception of videos~\cite{seshadrinathan2010motion}. However, the frame-level feature extraction paradigm inherently hinders the exploitation of spatio-temporal information for estimating the quality of videos.

In this work, we aim for dealing with the aforementioned limitations through model-based transfer learning strategies. Specifically, instead of leveraging pre-trained DNNs on object recognition~\cite{deng2009imagenet} for feature extraction, we propose to use human-annotated IQA databases to learn quality-aware frame-level feature representation. In addition, we employ a pre-trained 3D network~\cite{feichtenhofer2019slowfast} to capture the motion information. Two groups of features are delicately aggregated, leading to a complementary and effective spatio-temporal video representation. Moreover, we employ a mixed list-wise ranking loss function to train the entire BVQA model, which introduces additional performance gain. We summarize our contributions as follows:

\begin{itemize}
	
	\item[$\bullet$] We propose an effective and efficient method to learn a frame-level feature extractor for the VQA in the wild. We conduct a quality-aware pre-training on multiple IQA databases for transferring perceptually meaningful knowledge.
	
	\item[$\bullet$] We transfer the knowledge from an action recognition domain to perceive the motion distortion of videos. We empirically validate that the motion information is complementary to spatial features.

	\item[$\bullet$] We introduce a mixed list-wise ranking loss function for training the entire model, through which we obtain further performance improvement.
	
	\item[$\bullet$] Through extensive experiments, we verify that the proposed BVQA metric achieves the state-of-the-art (SOTA) results on six in-the-wild VQA databases.

\end{itemize}

\section{Related Work}\label{Sec:RelatedWork}

An intuitive solution to the BVQA task is applying a BIQA metric on videos frame by frame, followed by a features/scores pooling stage. In addition, motion information has also shown its promises in the perceived quality of videos~\cite{seshadrinathan2010motion, feichtenhofer2019slowfast, tu2020comparative}. Therefore, incorporating both spatial and motion information has become a promising paradigm for BVQA. We briefly review related BVQA methods following this line.

\subsection{Classical BVQA}\label{subsec:classical_bvqa}
A plethora of classical BVQA models relies on natural scene statistics (NSS), with an underlying assumption that the quality can be measured by the disturbance of NSS~\cite{yan2016blind}. NSS-based methods are derived from transform domains~\cite{moorthy2011blind, saad2012blind}, spatial domains~\cite{mittal2012no, mittal2013making}, or hybrid domains~\cite{ghadiyaram2017perceptual, ma2018blind}. Based on the 2D discrete-time transform (DCT) features of video frame-difference statistics, Saad \textit{et al.}~\cite{saad2014blind} further introduced motion information to enhance the representation capacity. Li \textit{et al.}~\cite{li2016spatiotemporal} captured the spatial and temporal regularities simultaneously using the 3D-DCT coefficients.  Mittal \textit{et al.}~\cite{mittal2016completely} designed a completely blind VQA metric by modeling the statistical naturalness of the videos and excavating the intersubband correlations. Dendi \textit{et al.}~\cite{dendi2020no} raised an asymmetric generalized Gaussian distribution (AGGD) to model the spatio-temporal statistics using 3D mean subtract contrast normalized coefficients and bandpass filter coefficients. Another line of work is the codebook-based methodology. Motivated by CORNIA~\cite{ye2012unsupervised}, Xu \textit{et al.}~\cite{xu2014no} proposed to learn the frame-level features via an unsupervised learning method and then used the support vector regression (SVR) to map feature representations to frame-level quality scores. The global video quality score is obtained using a temporal pooling.

\subsection{DNN-based BVQA}\label{subsec:dnn_bvqa}
In recent years, DNNs are inclined to dominate the BVQA field. Li \textit{et al.}~\cite{li2016no} extracted natural scene statistics using 3D shearlet transform and then made them more discriminative using a DNN, where a logistic regression function is used for training. Following an end-to-end learning framework MEON~\cite{Ma2018End}, Liu \textit{et al.}~\cite{liu2018end} devised a BVQA model that jointly optimizes the feature extractor, the codec classifier, and the quality predictor with a two-step training strategy. Zhang \textit{et al.}~\cite{zhang2019blind} pre-trained a DNN using the 3D-DCT coefficients with proxy labels. They then utilized a frequency histogram function to map the block-wise scores collected from the previous network to the perceptual quality. You \textit{et al.}~\cite{you2019deep} designed a BVQA model with a 3D convolutional neural network (CNN) as the feature extractor and a Long Short-Term Memory (LSTM) for the overall quality prediction. Li \textit{et al.} proposed a VSFA~\cite{li2019quality} model for quality assessment of in-the-wild videos, where two crucial effects of HVS, \textit{i.e.}, content-dependency and temporal-memory effects, are incorporated to account for quality-aware features. Based on VSFA, they then proposed a mixed databases training strategy towards a universal BVQA model (MDTVSFA)~\cite{li2021unified}. Ying \textit{et al.}~\cite{ying2021patch} created a local-to-global region-based BVQA architecture using a DNN that computes both 2D and 3D video features. Wang~\textit{et al.}~\cite{wang2021rich} aggregated several complementary 2D and 3D DNNs to incorporate different features for the BVQA task.

\section{Proposed Method}\label{Sec:ProposedMethod}

\begin{figure*}
  \centering
  \captionsetup{justification=centering}
  \includegraphics[width=1\textwidth]{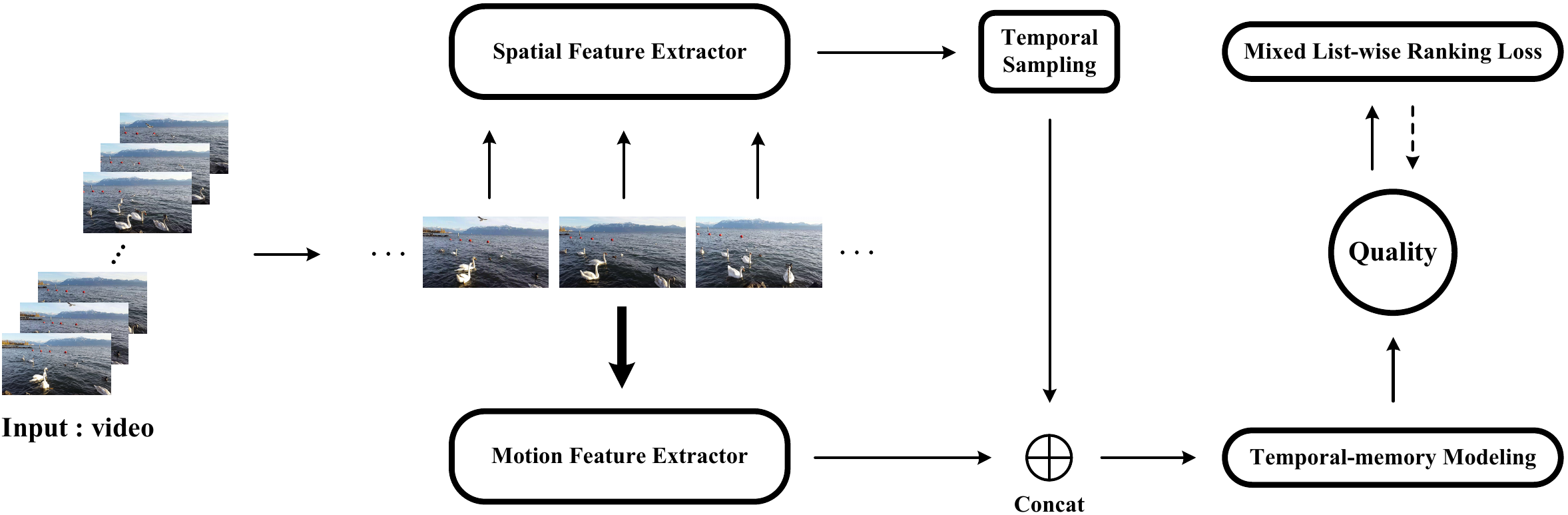}
  \caption{The overall framework of the proposed VQA model.}\label{Figure:OverallFramework}
\end{figure*}

In this section, we first describe the quality-aware pre-training strategy for learning the frame-level feature extractor. We then incorporate motion information to form a spatio-temporal representation, where special care is taken to fuse two groups of features in a reasonable way. Finally, we introduce a mixed list-wise ranking loss function to optimize the overall VQA model. The framework of our model is presented in Fig.~\ref{Figure:OverallFramework}.

\subsection{Quality-aware Pre-training}\label{Subsec:QualityAwarePrT}
\subsubsection{Transfer Learning}\label{Subsubsec:TransferLearning}
The lack of large video databases with human quality annotations is a common obstacle to applying DNNs for BVQA. As a consequence, the pre-training followed by fine-tuning is a widely-used transfer learning paradigm due to its favorable flexibility that poses no constraint on the label spaces of the source and target domains~\cite{pan2010survey}, for which we have a formulation~\cite{tlbook}:
\begin{equation}\label{Equation:TransferLearning}
f_{s}^{*} = \argmin_{f_{s} \in \mathcal{H}} \frac{1}{N_{s}} \sum_{i=1}^{N_{s}}\ell_{s}\left(f_{s}(x_{s,i}, q_{s,i})\right) + \alpha_{s} R\left(D_{t},f_{s})\right)
\end{equation}
where $(x_{s,i},q_{s,i})$ is the $i$-th tuple of the sample and label in the source domain, $N_{s}$ is the number of samples in the source domain, $f_{s}$ is a function that lies in a Hilbert space $\mathcal{H}$, which we optimize with the loss function $\ell_{s}$ using the data of source domain $D_{s}$, and $R(\cdot)$ is a regularization term controlled by a weight $\alpha_{s}$, whose objective is leveraging or finetuning the $f_{s}$ in the target domain $D_{t}$. Although being popular in VQA~\cite{li2019quality, li2021unified, tu2021rapique}, the effectiveness of this paradigm is limited by distributional shifts between the source (object recognition) and target domains (VQA).

\subsubsection{Source Domain Selection}\label{Subsubsec:SourceDomainSelection}
Considering that videos are composed of multiple stacked images (frames), we aim for transferring the knowledge from the IQA databases with authentic distortions, which we assume to be source domains that better match the target domains, \textit{i.e.}, VQA in the wild.

To verify the rationality of the selected source domains, we quantify the distances between the source and target domains. Specifically, we use the CORAL~\cite{sun2016deep} as a proxy metric to measure the feature distance between the source and target domains. For the source domain, we take one image classification database (\textit{i.e.}, ImageNet~\cite{deng2009imagenet}) and four IQA databases (\textit{i.e.}, BID~\cite{ciancio2010no}, LIVE Challenge~\cite{ghadiyaram2016massive}, KonIQ-10k~\cite{hosu2020koniq}, and SPAQ~\cite{fang2020perceptual}) for comparison. Here, we uniformly sample the same number of images from ImageNet~\cite{deng2009imagenet} across all semantic categories. As for the target domains, we acquire videos from five in-the-wild VQA databases, \textit{i.e.}, CVD2014~\cite{nuutinen2016cvd2014}, KoNViD-1k~\cite{hosu2017konstanz}, LIVE-Qualcomm~\cite{ghadiyaram2018capture}, LIVE-VQC~\cite{sinno2019large}, and YouTube-UGC~\cite{wang2019youtube}. We then use the pre-trained ResNet-50~\cite{he2016deep} on ImageNet to extract features of all samples in the source and target domains. Note that the target features of videos are obtained by the average pooling of their frame-level features. The pairwise CORAL distances are shown in Table~\ref{Table:CoralDistance} in Section~\ref{subsubsec:AblationStudy}, from which we observe that IQA databases are statistically closer to the target VQA domain compared with ImageNet. Besides, we find that no single IQA database is consistently to be the closest domain with different VQA databases. Thus, it is highly desirable to specify an effective learning scheme to transfer knowledge from multiple diverse IQA databases, which can cover a broad range of appropriate content and authentic distortions.


\subsubsection{Training Frame-level Feature Extractor}\label{Subsubsec:QualityAwareLearning} Inspired by the database combination method~\cite{zhang2021uncertainty}, we leverage multiple IQA databases for pre-training the function $f_{s}$, which will serve as a frame-level feature extractor. We first formulate the loss function $\ell_{s}$ to make full use of the training data. Given an image pair $(x_{s}, y_{s})$ sampled from an IQA database, under the Thurstone's model~\cite{thurstone1927law}, their perceptual quality $s(x_{s})$ and $s(y_{s})$ are assumed to follow Gaussian distributions with means $(\mu(x_{s}), \mu(y_{s}))$ and standard deviations (std) $(\sigma(x_{s}),\sigma(y_{s}))$, respectively. Assuming the variability of quality across images is uncorrelated, their quality difference $s(x_{s})-s(y_{s})$ also conforms to a Gaussian distribution with mean $\mu(x_{s})-\mu(y_{s})$ and std $\sqrt{\sigma^2(x_{s})+\sigma^2(y_{s})}$. Through a frame-level quality prediction network $f_{s}=\{\phi, h_{\mu}, h_{\sigma}\}$ parameterized by a vector $\bm{w}$, where $\phi$ and $h_{\mu}$ / $h_{\sigma}$ denote the backbone network and the fully-connected (FC) layers, the estimated mean and std can be computed as $\mu_{\bm{w}}(\cdot) = h_{\mu}(\phi(\cdot))$ and $\sigma_{\bm{w}}(\cdot) = h_{\sigma}(\phi(\cdot))$, respectively. The estimated quality difference is also assumed to follow a Gaussian distribution with mean $\mu_{\bm{w}}(x_{s})-\mu_{\bm{w}}(y_{s})$ and std $\sqrt{\sigma_{\bm{w}}^2(x_{s})+\sigma_{\bm{w}}^2(y_{s})}$, for which we simultaneously supervise the learning of mean and std. The probabilities (Pr) that $x_{s}$ is of higher perceptual quality than $y_{s}$ according to the ground truths and the predicted scores are as follows:
\begin{equation}\label{Equation:Prob}
\begin{split}
p(x_{s},y_{s})={\rm{Pr}}(s(x_{s}) \ge s(y_{s})) \\
=\Phi \left( {\frac{\mu(x_{s})-\mu(y_{s})}{\sqrt{\sigma^2(x_{s})+\sigma^2(y_{s})}}} \right)
\end{split}
\end{equation}

\begin{equation}\label{Equation:ProbPredict}
\begin{split}
p_{\bm{w}}(x_{s},y_{s})={\rm{Pr}}(s_{\bm{w}}(x_{s}) \ge s_{\bm{w}}(y_{s})) \\
=\Phi \left( {\frac{\mu_{\bm{w}}(x_{s})-\mu_{\bm{w}}(y_{s})}{\sqrt{\sigma_{\bm{w}}^2(x_{s})+\sigma_{\bm{w}}^2(y_{s})}}} \right)
\end{split}
\end{equation}
where $\Phi(\cdot)$ denotes the Gaussian cumulative distribution function. Note that when scaling $\mu_w(\cdot)\rightarrow \rho \mu_w(\cdot)$ and $\sigma_w(\cdot) \rightarrow \rho\sigma_w(\cdot)$, the probability $p_w(\cdot,\cdot)$ inferred by Eq.~\eqref{Equation:ProbPredict} is unchanged. To avoid this scaling ambiguity and supply $\sigma_{\bm{w}}(\cdot)$ with a direct supervision, we enforce a regularizer of $\sigma_{\bm{w}}(\cdot)$ for std learning. For an image pair $(x_{s}, y_{s})$, a binary label $g$ is assigned as $g(x_{s},y_{s})=\rm{sign}(\sigma(x_{s})-\sigma(y_{s}))$.
Empirically, the similarity of Gaussian distribution and the uncertainty of the regularizer can be measured by the fidelity loss~\cite{tsai2007frank} and the hinge loss respectively as follows:
\begin{equation}\label{Equation:LossP}
\begin{split}
\ell_P\{(x_{s},y_{s}),p;{\bm{w}}\}=1-\sqrt{p(x_{s},y_{s}) \cdot p_{\bm{w}}(x_{s},y_{s})} \\
-\sqrt{(1-p(x_{s},y_{s})) \cdot (1-p_{\bm{w}}(x,y_{s}))}
\end{split}
\end{equation}

\begin{equation}\label{Equation:LossG}
\begin{split}
& \ell_G\{(x_{s},y_{s}),g;{\bm{w}}\}=\\
& {\rm{max}}(0,\eta-{\rm{sign}}(\sigma(x_{s})-\sigma(y_{s})) \cdot (\sigma_{\bm{w}}(x_{s})-\sigma_{\bm{w}}(y_{s})))
\end{split}
\end{equation}
where $\eta$ is a margin constant. In practice, we randomly sample a large number of image pairs from the aforementioned four IQA databases, resulting in a set $\bm{\Omega} = \{(x_{s}, y_{s})_i, p_i, g_i\}_{i=1}^{N_{s}}$ for training. At the training stage, we utilize every batch $\mathcal{B}$ to optimize ${\bm{w}}$ using the overall loss:
\begin{equation}\label{Equation:LossIQA}
\ell_{s}(\mathcal{B};{\bm{w}})=\frac{1}{|\mathcal{B}|}\sum\limits_{\mathcal{B} \in \bm{\Omega}} \ell_P\{(x_{s},y_{s}),p;{\bm{w}}\}+\nu\ell_G\{(x_{s},y_{s}),g;{\bm{w}}\}
\end{equation}
where $\nu$ is a balance coefficient. In practice, we use a variant of stochastic gradient descent (SGD) algorithm with a $L_{2}$ weight decay as the regularizer $R$ to optimize the network. Once the training is completed, we extract the frame-level features of videos using the backbone network $\phi$.

\begin{figure*}[htbp]
  \centering
  \captionsetup{justification=centering}
  \includegraphics[width=0.94\textwidth]{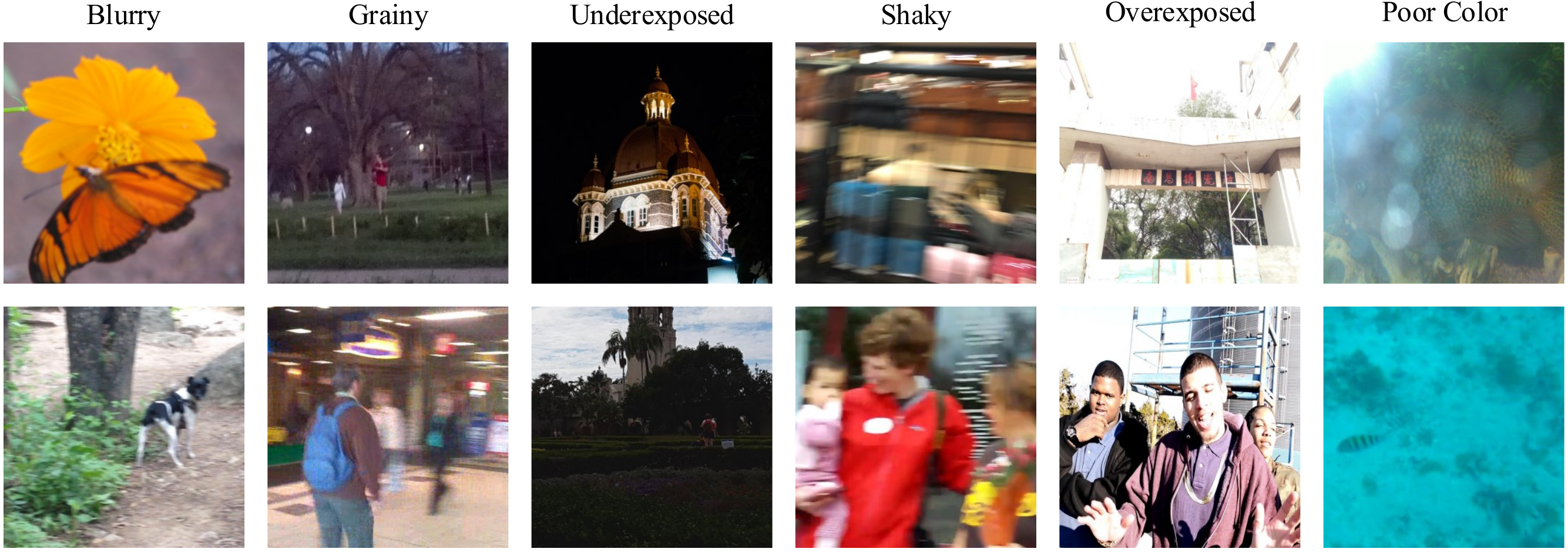}
  \caption{Comparison with images from IQA databases and single frames of videos from VQA databases. The top row presents the images from IQA databases including BID~\cite{ciancio2010no}, LIVE Challenge~\cite{ghadiyaram2016massive}, KonIQ-10k~\cite{hosu2020koniq}, and SPAQ~\cite{fang2020perceptual}. And the bottom row presents the frames of videos sampled from VQA databases covering CVD2014~\cite{nuutinen2016cvd2014}, KoNViD-1k~\cite{hosu2017konstanz}, LIVE-Qualcomm~\cite{ghadiyaram2018capture}, LIVE-VQC~\cite{sinno2019large}, YouTube-UGC~\cite{wang2019youtube}, and LSVQ~\cite{ying2021patch}.}\label{Figure:IQARelatedVideoDistortion}
\end{figure*}

\subsubsection{Qualitative Demonstration}\label{Subsubsec:QualitativeDemonstration}
To verify the rationality of the quality-aware pre-training more intuitively, we present some visual examples with representative types of realistic impairments between the source and target domains, which including ``Blurry'', ``Grainy'', ``Underexposed'', ``Shaky'', ``Overexposed'', and ``Poor Color'' as shown in Fig.~\ref{Figure:IQARelatedVideoDistortion}. Each sample is labeled with a single dominant distortion for better visualization. From Fig.~\ref{Figure:IQARelatedVideoDistortion}, we can observe similar distortion patterns between images sampled from public IQA databases and single frames from videos on VQA databases.

\subsection{Motion Perception}\label{subsec:motion}
In addition to spatial appearance, dynamic changes are deemed as the most distinctive characteristic of videos~\cite{kwon2020motionsqueeze}. A plethora of biological researches on the primate visual structure~\cite{livingstone1988segregation, felleman1991distributed, van1994neural} demonstrated that there are approximately 15-20\% M-cells sensitive to fast temporal changes. Therefore, incorporating motion information is helpful to facilitate video quality estimation. Previous work captured motion information using various hand-crafted features such as silhouette~\cite{blank2005actions} and optical flow~\cite{dalal2006human}. These methods are either computationally expensive or with less representational power. We resort to a learning-based method for extracting motion features. Similar to the model-based transfer learning philosophy stated in Section~\ref{Subsec:QualityAwarePrT}, we make use of a pre-trained 3D-DNN on the action recognition to extract motion features of videos. Specifically, we resort to the fast pathway of the pre-trained SlowFast (dubbed as SlowFast$_F$) network~\cite{feichtenhofer2019slowfast} on Kinetics-400~\cite{kay2017kinetics}, which contains rich motion-related contents. SlowFast$_F$ can produce motion features with high temporal resolution since it maintains temporal fidelity as much as possible by prohibiting temporal downsampling before the last pooling layer. Besides, SlowFast$_F$ is formed in a lightweight manner with low channel capacity, which makes it more computationally efficient. As a result, the extracted features are sensitive to fast motion, which is complementary to the spatial features.

\begin{figure*}[th]
    \centering
    \captionsetup{justification=centering}
    \subfloat[Breakdancing]{\includegraphics[width=0.45\textwidth]{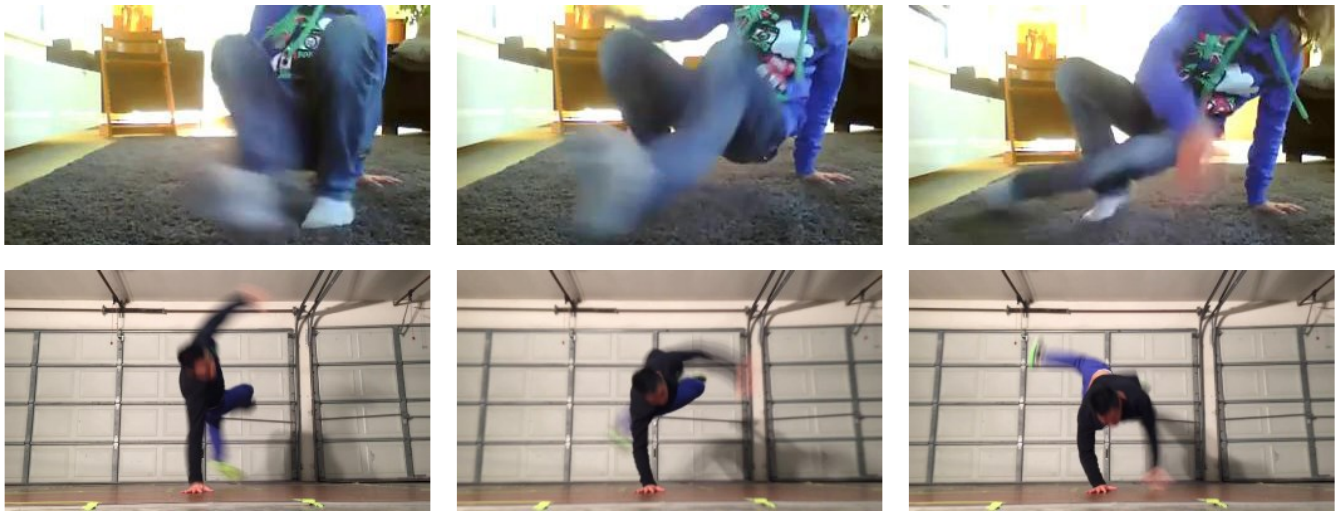}}\hskip2em
    \subfloat[Riding with horse]{\includegraphics[width=0.45\textwidth]{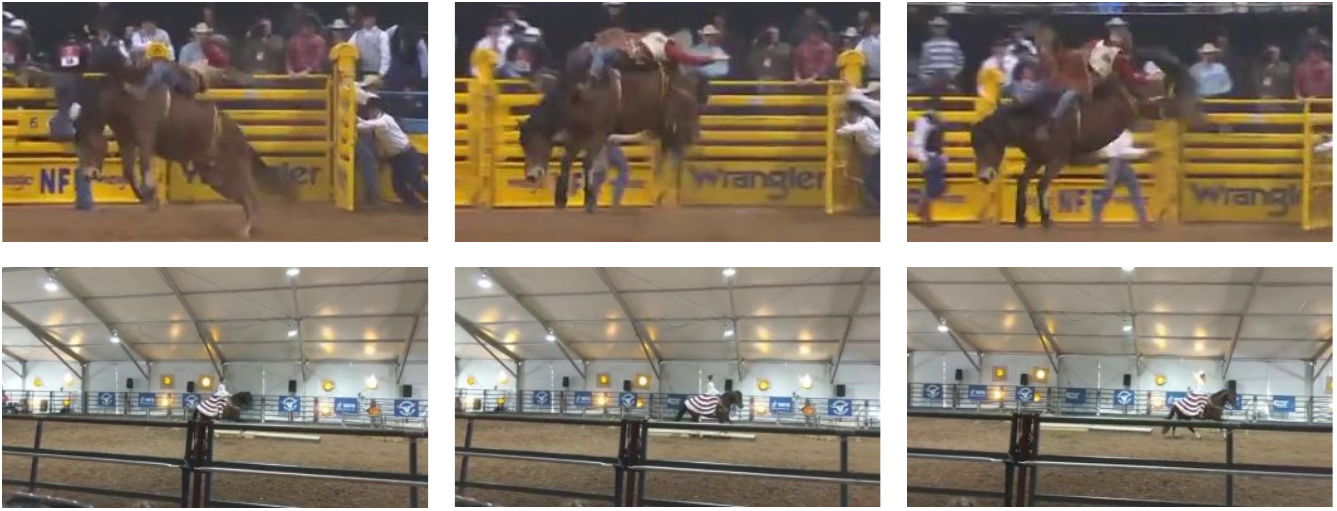}}
    \vspace{2pt}
    \subfloat[Cooking on campfire]{\includegraphics[width=0.45\textwidth]{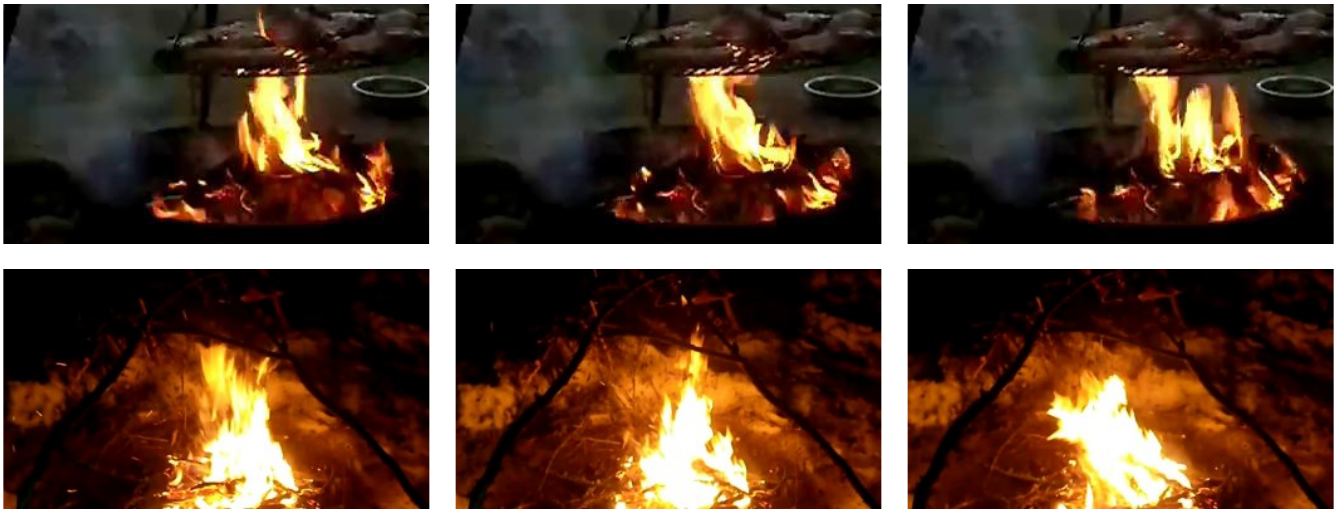}}\hskip2em
    \subfloat[Playing basketball]{\includegraphics[width=0.45\textwidth]{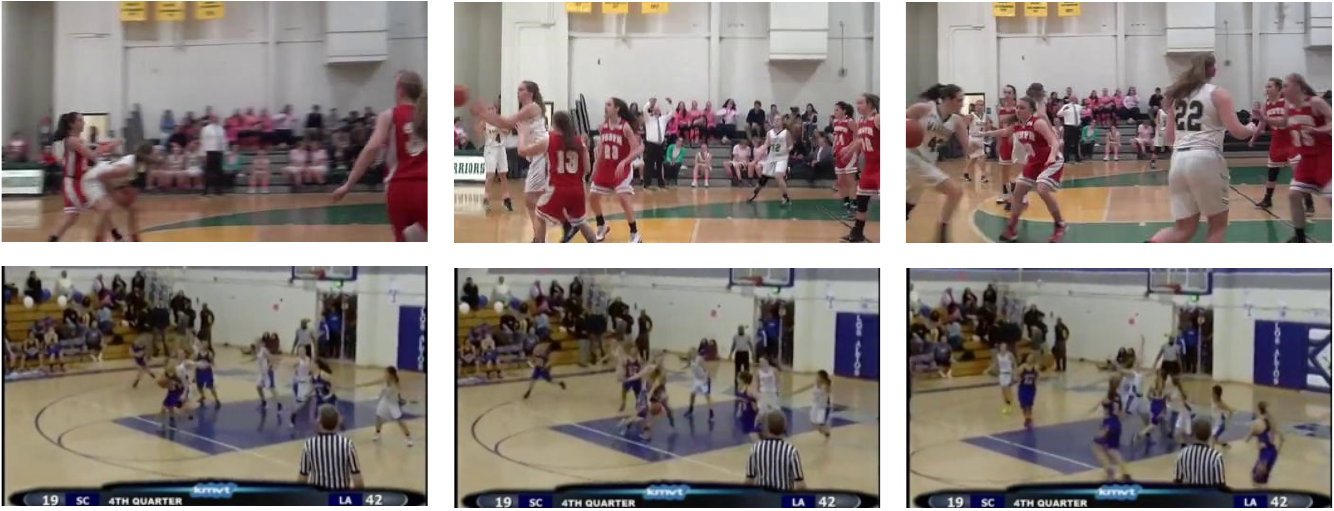}}
    \vspace{2pt}
    \caption{Comparison of videos with different motion patterns. In each subgraph, the top row presents the continuous frames of videos sampled from Kinetics-400~\cite{kay2017kinetics}, while the bottom row presents that from VQA databases including KoNViD-1k~\cite{hosu2017konstanz}, LIVE-VQC~\cite{sinno2019large}, and LSVQ~\cite{ying2021patch}.}\label{Figure:MotionRelatedVideoDistortion}
\end{figure*}


As shown in Fig.~\ref{Figure:MotionRelatedVideoDistortion}, we exhibit continuous frames of videos sampled from Kinetics-400 and that from VQA databases, from which we observe a similar distortion pattern. Specifically, all the compared videos undergo significant motion blur distortion regardless of their contents.

\subsection{Features Fusion}\label{subsection:FeaturesFusion}
Let $\{T, H \times W, C\}$ denote the temporal, spatial, and channel dimensions of a raw video clip $\bm{z} = \{z_{t}\}_{t=1}^{T}$ where $z_{t}$ is the $t$-th frame of the video. We use the proposed quality-aware pre-training scheme to train a ResNet-50~\cite{he2016deep} as the frame-level feature extractor $\phi$. The network architecture is briefly summarized in ``Spatial'' column of Table~\ref{Table:NetworkArchitecture}. To aggregate the spatial information, we leverage the activation of the last convolution of the ``Spatial" pipeline. To obtain a rich spatial feature representation, we use both the global average pooling (GAP) and the global standard deviation pooling (GSP) to aggregate spatial features of a single frame as $v_{s,t} =  \rm{GAP}(\phi(z_{t})) \oplus \rm{GSP}(\phi(z_{t}))$, where $\oplus$ denotes the concatenation operation and $\phi(z_{t})$ is with the size of $\{T, H/32 \times W/32, 2048\}$. As a result, we can obtain 4,096-dimensional feature vectors with a temporal length of $T$.

In the SlowFast$_F$ pipeline, we use the default parameters as described in~\cite{feichtenhofer2019slowfast} where the temporal stride in the slow pathway is $\tau=8$, the speed and channel ratios in the fast pathway are $\alpha=4$ and $\beta=1/8$, respectively. The network architecture is briefly summarized in the ``Motion'' column of Table~\ref{Table:NetworkArchitecture}. Given the video clip $\bm{z}$, SlowFast$_F$ can produce an activation with a size of $\{T/2, H/32 \times W/32, 256\}$. Similarly, the features are spatially pooled using GAP and GSP, resulting in a sequence of 512-dimensional frame-level features $\{v_{m, t}\}_{t=1}^{T/2}$.

It remains to fuse the spatial and motion features in a reasonable manner. To match the temporal resolution of the motion pipeline, we temporally sample one out of every two frames of the spatial feature tensor, resulting in a 4,096-dimensional tensor with a length of $T/2$. We then concatenate the spatial and motion features along the channel dimension as $v_{t} = v_{s, t} \oplus v_{m,t}$. Finally, we have 4,608-dimensional frame-level feature vectors with a temporal length of $T/2$.

\newcommand{\blockc}[4]{\multirow{#4}{*}{\(\left[\begin{array}{c}\text{1$\times$1$^2$, #1}\\[-.1em] \text{1$\times$3$^2$, #1}\\[-.1em] \text{1$\times$1$^2$, #2} \end{array}\right]\)$\times$#3}
}

\newcommand{\blockv}[4]{\multirow{#4}{*}{\(\left[\begin{array}{c}\text{3$\times$1$^2$, #1}\\[-.1em] \text{1$\times$3$^2$, #1}\\[-.1em] \text{1$\times$1$^2$, #2} \end{array}\right]\)$\times$#3}
}

\newcommand{\blocko}[2]{\multirow{#2}{*}{\(\begin{array}{c}\text{S: $T\times$$\frac{H}{#1}$$\times$$\frac{W}{#1}$}\\[0.2em] \text{M: $\frac{T}{2}$$\times$$\frac{H}{#1}$$\times$$\frac{W}{#1}$} \end{array}\)}
}

\newcommand{\blockoa}[1]{\multirow{#1}{*}{\(\begin{array}{c}\text{S: $T\times H\times W$}\\[0.2em] \text{M: $\frac{T}{2}$$\times H\times W$} \end{array}\)}
}

\newcommand{\blockob}[1]{\multirow{#1}{*}{\(\begin{array}{c}\text{S: $T\times1\times1$}\\[0.2em] \text{M: $\frac{T}{2}\times1\times1$} \end{array}\)}
}

\newcommand{\blockoc}[1]{\multirow{#1}{*}{\(\begin{array}{c}\text{S: $T\times2$}\\[0.2em] \text{M: \#classes} \end{array}\)}
}

\newcommand{\blockconv}[3]{\multirow{#3}{*}{\(\begin{array}{c}\text{#1$\times$7$^2$, #2}\\[0.2em] \text{stride 1, 2$^2$} \end{array}\)}
}

\newcommand{\blockpool}[1]{\multirow{#1}{*}{\(\begin{array}{c}\text{1$\times$3$^2$ max}\\[0.2em] \text{stride 1, 2$^2$} \end{array}\)}
}

\begin{table}[t]
	\centering
	\caption{The architectures of the ``Spatial'' and ``Motion'' sub-networks. The kernel dimensions are denoted by \{T$\times$(H$\times$W),C\} for temporal, spatial, and channel dimensions. Strides are denoted as \{temporal stride, spatial stride$^2$\}}
    \scalebox{0.85}{
	\begin{tabular}{l |c |c |c}
		\toprule
		{\bf Layer} & {\bf Spatial} & {\bf Motion} & {\bf Output size} \\
		\hline
        {Input} & --- & --- & {$T\times H \times W$} \\
        \hline
        \multirow{3}{*}{data layer} & \multirow{3}{*}{---} & \multirow{3}{*}{stride 2, 1$^2$} & \blockoa{3} \\
        & & & \\
        & & & \\
        \hline
		\multirow{3}{*}{Conv} & \blockconv{1}{64}{3} & \blockconv{5}{8}{3} & \blocko{2}{3} \\
        & & & \\
        & & & \\
		\hline	
		\multirow{3}{*}{Pooling} & \blockpool{3} & \blockpool{3} & \blocko{4}{3} \\
        & & & \\
        & & & \\
		\hline
        \multirow{4}{*}{ResB 1}& \blockc{64}{256}{3}{4} & \blockv{8}{32}{3}{4} & \blocko{4}{4} \\
        & & & \\
        & & & \\
        & & & \\
		\hline
        \multirow{4}{*}{ResB 2}& \blockc{128}{512}{4}{4} & \blockv{16}{64}{4}{4} & \blocko{8}{4} \\
        & & & \\
        & & & \\
        & & & \\
        \hline
       \multirow{4}{*}{ResB 3}& \blockc{256}{1024}{6}{4} & \blockv{32}{128}{6}{4} & \blocko{16}{4} \\
        & & & \\
        & & & \\
        & & & \\
        \hline
        \multirow{4}{*}{ResB 4}& \blockc{512}{2048}{3}{4} & \blockv{64}{256}{3}{4} & \blocko{32}{4} \\
        & & & \\
        & & & \\
        & & & \\
        \hline
	    \multirow{3}{*}{Pooling} & \multirow{3}{*}{average} & \multirow{3}{*}{average} & \blockob{3} \\
        & & & \\
        & & & \\
        \hline
		\multirow{3}{*}{FC} & \multirow{3}{*}{(2,048)$\times$2} & \multirow{3}{*}{256} & \blockoc{3} \\
        & & & \\
        & & & \\
		\bottomrule
	\end{tabular}}
	\label{Table:NetworkArchitecture}
\end{table}

\subsection{Temporal Modeling and Quality Prediction}
Similar to~\cite{li2021unified}, we take the temporal-memory effect into consideration. Specifically, we use a gated recurrent unit~\cite{cho2014learning} (GRU) to model the temporal information. To enable efficient learning of the GRU, a dimension reduction is performed to the frame-level feature vectors $v_t$ using a FC layer:
\begin{equation}\label{Equation:DimReduct}
\hat{v}_t = \mathcal{W}_{v}v_{t} + b_{v}.
\end{equation}
where $\mathcal{W}_{v}$ and $b_{v}$ are learnable parameters of the dimension reduction FC layer. Given $\{\hat{v}_t|t=1,2,...,T\}$ as the input to GRU, the hidden state at the $t$-th time step $h_t$ depends on both the previous state $h_{t-1}$ and the current input feature $\hat{v}_t$ as:
\begin{equation}\label{Equation:GRU}
h_t={\rm{GRU}}(\hat{v}_t, h_{t-1})
\end{equation}
We then use an FC layer to map the sequence of hidden states to the frame-level quality scores $\{q_t\}_{t=1}^{T}$.

We then adopt the HVS-inspired temporal hysteresis pooling~\cite{seshadrinathan2011temporal} to temporally aggregate the frame-level quality scores to an overall video quality score. Specifically, we use a differentiable hysteresis-based temporal pooling model. Let $\tau$ denotes the memory duration, a memory quality item $m_t$ at the t-th frame is defined with the worst quality case across the previous frames as:
\begin{equation}\label{Equation:MemQuality}
m_t=\left\{
\begin{array}{ll}
q_t, & {t=1} \\
{\rm{min}}({q_k}) \quad k\in\{{\rm{max}}(1,t-\tau),...,t-1\}, & {t>1}
\end{array}
\right.
\end{equation}
A current quality item $c_t$ is calculated with the next $\tau$ frames based on the fact that more rapid response will be paid into the drops in quality than the increase situation. This procedure can be established by a weighted quality combination using the softmin function as:
\begin{equation}\label{Equation:CurQuality}
\begin{aligned}
\begin{gathered}
c_t=\sum\nolimits_{k}a_kq_k \\
a_k=\left. e^{-q_k} \middle/ \sum\nolimits_{i}e^{-q_i} \right. \\
i,k \in \{t,..., {\rm{min}}(t+\tau,T)\}
\end{gathered}
\end{aligned}
\end{equation}
Then the hysteresis effect is expressed by a linear combination of the memory and the current quality items as:
\begin{equation}\label{Equation:TPooling}
q_t^{\prime}=\beta m_t + (1-\beta) c_t
\end{equation}
where $\beta$ is a contribution factor of different components.
Finally, the entire video quality score is computed as the global average of the time-varying predicted scores:
\begin{equation}\label{Equation:FianlQuality}
Q_p=\frac{1}{T} \sum_{t=1}^T q_t^{\prime}
\end{equation}

\subsection{Loss Function}
\label{Sec:LossFunction}

An objective video quality model is expected to make quality predictions of videos consistently with subjective ratings. To this end, we employ two loss functions to encourage prediction monotonicity and precision, respectively. Following~\cite{liu2018end, li2021unified}, we adopt the Pearson Linear Correlation Coefficient (PLCC) loss to optimize our model towards higher prediction precision. To better measure the degree of linear correlation against ground truths, a nonlinear mapping is commonly introduced before calculating PLCC \cite{zhang2020blind, chen2020perceptual}. Following the recommendation of the Video Quality Experts Group, this procedure can be implemented with a 4-parameter logistic function \cite{video2000final} as:
\begin{equation}\label{Equation:NonlinearMapping_Standard}
Q_m=\frac{\gamma_3^{\prime}-\gamma_4^{\prime}}{1+e^{-\frac{Q_p-\gamma_1^{\prime}}{|\gamma_2^{\prime}|}}} + \gamma_4^{\prime}
\end{equation}
where $\{ \gamma_i^{\prime} | i \in \{1,...,4\} \}$ are the learnable fitting parameters and $Q_m$ is the fitted quality score. As the reformulation in~\cite{li2021unified}, the above 4-parameter logistic function can be designed as a network module of \{Linear, Sigmoid, and Linear\} layers, which is represented as:
\begin{equation}\label{Equation:NonlinearMapping}
Q_m=\gamma_3 {\rm{Sigmoid}} (\gamma_1 Q_p + \gamma_2) + \gamma_4
\end{equation}
where $\gamma_1=1/|\gamma_2^{\prime}|$, $\gamma_2=-\gamma_1^{\prime}/|\gamma_2^{\prime}|$, $\gamma_3=\gamma_3^{\prime}-\gamma_4^{\prime}$, $\gamma_4=\gamma_4^{\prime}$, and ${\rm{Sigmoid(\cdot)}}=1/\left(1+e^{-(\cdot)}\right)$. Given $N$ training samples from a specific database, the differentiable PLCC loss then can be formulated as:
\begin{equation}\label{Equation:LossPLCC}
\begin{aligned}
\begin{gathered}
\ell_{\rm{PLCC}}=(1-{\rm{PLCC}})/2 \\
{\rm{PLCC}}=\frac{\sum\nolimits_i (Q_m^i-\overline{Q}_m)(Q^i-\overline{Q})}{\sqrt{\sum\nolimits_i (Q_m^i-\overline{Q}_m)^2 \sum\nolimits_i (Q^i-\overline{Q})^2}}
\end{gathered}
\end{aligned}
\end{equation}
where $\overline{Q}_m$ and $\overline{Q}$ denote the mean values of the fitted predictions $\{Q_m^{i}\}_{i=1}^{N}$ and subjective quality opinions $\{Q^{i}\}_{i=1}^{N}$.

To the best of our knowledge, existing BVQA methods have not explored any optimization strategy to explicitly encourage the prediction monotonicity of models. This is mainly due to the non-differentiable operations of frequently used order statistics and ranking metrics. Inspired by~\cite{blondel2020fast}, we adopt a differentiable proxy to boost the model prediction monotonicity, which is termed as a Spearman Rank-order Correlation Coefficient (SRCC) loss. In principle, the SRCC metric can be defined as the PLCC between ranks. We denote the ranks of the model predictions $\{Q_{p}^{i}\}_{i=1}^{N}$ and the ground-truth annotations $\{Q^{i}\}_{i=1}^{N}$ as $\{Q_{pr}^{i}\}_{i=1}^{N}$ and $\{Q_{r}^{i}\}_{i=1}^{N}$ respectively, where we assume the elements of the original model predictions are ranked in a descending order. The differentiable SRCC loss can be formulated as:
\begin{equation}\label{Eq4:LossSRCC}
\begin{aligned}
\begin{gathered}
\ell_{\rm{SRCC}}=1-{\rm{SRCC}} \\
{\rm{SRCC}}=\frac{\sum\nolimits_i (Q_{pr}^i-\overline{Q}_{pr})(Q_r^i-\overline{Q_r})}{\sqrt{\sum\nolimits_i (Q_{pr}^i-\overline{Q}_{pr})^2 \sum\nolimits_i (Q_{r}^i-\overline{Q}_{r})^2}}
\end{gathered}
\end{aligned}
\end{equation}
where ${Q}_{pr}$ and ${Q}_{r}$ are computed from a differentiable ranking function. We refer readers to~\cite{blondel2020fast} for details of the process of computing "soft" ranks and the proof of its differentiability. Finally, we have an overall loss function as:
\begin{equation}\label{Equation:LossVQA}
\ell = \ell_{\rm{PLCC}} + \lambda \ell_{\rm{SRCC}}
\end{equation}
where $\lambda$ trades off the influence of the two elements. Note that both SRCC loss and PLCC loss are list-wise ranking loss functions, which can be used in either individual database training or mixed databases training~\cite{li2021unified} settings.


\section{Experiments}\label{Sec:Experiments}
In this section, we first describe the experimental setups, including benchmarking databases, competing methods, performance criteria, and implementation details. We then present and analyze the results of three scenarios: individual, mixed, and cross databases. Finally, we verify the rationality of the proposed method through qualitative results, ablation study, and computational complexity analysis.
\subsection{Experimental Setups}\label{subsec:expsetup}
\subsubsection{Benchmarking Databases}\label{subsec:VQAdatabase}

\begin{table*}[!htbp]
  \centering
  \caption{Summary of the benchmarking in-the-wild VQA databases}\label{Table:VQAdatabase}
  \begin{threeparttable}
  \begin{tabular}{l|ccccccc}
    \toprule
        {\bf Database} & Number of Videos & Number of Scenes & Resolution & Format & Time Duration & Max Length & Annotation Range \\
    \hline
        CVD2014~\cite{nuutinen2016cvd2014}         &    234  &      5 & 480p, 720p & RGB & 10-25s & 830 & [-6.50, 93.38] \\
        KoNViD-1k~\cite{hosu2017konstanz}          &  1,200  &  1,200 & 540p & RGB & 8s & 240 & [1.22, 4.64] \\
        LIVE-Qualcomm~\cite{ghadiyaram2018capture} &    208  &     54 & 1080p & YUV & 15s & 526 & [16.5621, 73.6428] \\
        LIVE-VQC~\cite{sinno2019large}             &    585  &    585 & 240p-1080p & RGB & 10s & 1,202 & [6.2237, 94.2865] \\
        YouTube-UGC~\cite{wang2019youtube}         &  1,142\tnote{*}  &  1,142 & 360p-4k & YUV & 20s & 2,819 & [1.242, 4.698] \\
        LSVQ~\cite{ying2021patch}                  & 39,072\tnote{**}  & 39,072 &  99p-4k & RGB & 5-12s & 4,096 & [2.4483, 91.4194]\\
    \bottomrule
  \end{tabular}
  \begin{tablenotes}
        \footnotesize
        \item[*] Following~\cite{wang2019youtube}, we exclude 57 grayscale videos in latest YouTube-UGC database, remaining 1,142 videos.
        \item[**] The number of existing ground truths (MOSs) in LSVQ is 39,072.
  \end{tablenotes}
  \end{threeparttable}
\end{table*}

\begin{table*}[t]
  \footnotesize
  \centering
  \caption{Median SRCC and PLCC results on CVD2014, KoNViD-1k, LIVE-Qualcomm, LIVE-VQC, and YouTube-UGC under the individual database training setting. The standard deviation is shown in grey}
  \begin{tabular}{l|l|cccccc}
      \toprule
    \multirow{2}{*} {} & \bf{Database} & {CVD2014} & {KoNViD-1k} & {LIVE-Qualcomm} & {LIVE-VQC} & {YouTube-UGC} & {W.A.} \\
    \cline{2-8}
    & \bf{Criteria} & \multicolumn{5}{c}{SRCC} \\
    \hline
    \multirow{10}{*} {BIQA} & NIQE & 0.4755 ({\color{grey}$\pm$ 0.1174}) & 0.5392 ({\color{grey}$\pm$ 0.0366}) & 0.4608 ({\color{grey}$\pm$ 0.1049}) & 0.5930 ({\color{grey}$\pm$ 0.0581}) & 0.2499 ({\color{grey}$\pm$ 0.0532}) & 0.4412 \\
       & IL-NIQE & 0.5295 ({\color{grey}$\pm$ 0.1039}) & 0.5199 ({\color{grey}$\pm$ 0.0377}) & 0.0556 ({\color{grey}$\pm$ 0.1489}) & 0.5019 ({\color{grey}$\pm$ 0.0669}) & 0.3198 ({\color{grey}$\pm$ 0.0468}) & 0.4209 \\
       & BRISQUE & 0.7900 ({\color{grey}$\pm$ 0.0570}) & 0.6493 ({\color{grey}$\pm$ 0.0416}) & 0.5527 ({\color{grey}$\pm$ 0.1029}) & 0.5936 ({\color{grey}$\pm$ 0.0634}) & 0.3932 ({\color{grey}$\pm$ 0.0613}) & 0.5566 \\
       & M3 & 0.8009 ({\color{grey}$\pm$ 0.0482}) & 0.6422 ({\color{grey}$\pm$ 0.0408}) & 0.6272 ({\color{grey}$\pm$ 0.0948}) & 0.5876 ({\color{grey}$\pm$ 0.0649}) & 0.3450 ({\color{grey}$\pm$ 0.0528}) & 0.5421 \\
       & HIGRADE & 0.7096 ({\color{grey}$\pm$ 0.0768}) & 0.7062 ({\color{grey}$\pm$ 0.0333}) & 0.6326 ({\color{grey}$\pm$ 0.0843}) & 0.5959 ({\color{grey}$\pm$ 0.0653}) & 0.7252 ({\color{grey}$\pm$ 0.0340}) & 0.6892 \\
       & FRIQUEE & 0.8212 ({\color{grey}$\pm$ 0.0430}) & 0.7352 ({\color{grey}$\pm$ 0.0275}) & 0.7158 ({\color{grey}$\pm$ 0.0842}) & 0.6502 ({\color{grey}$\pm$ 0.0553}) & 0.7538 ({\color{grey}$\pm$ 0.0286}) & 0.7315 \\
       & CORNIA & 0.6277 ({\color{grey}$\pm$ 0.0846}) & 0.7351 ({\color{grey}$\pm$ 0.0274}) & 0.4551 ({\color{grey}$\pm$ 0.1155}) & 0.6808 ({\color{grey}$\pm$ 0.0460}) & 0.5671 ({\color{grey}$\pm$ 0.0422}) & 0.6440 \\
       & HOSA & 0.8478 ({\color{grey}\bf{$\pm$ 0.0349}}) & 0.7606 ({\color{grey}$\pm$ 0.0239}) & 0.7300 ({\color{grey}$\pm$ 0.0793}) & 0.6784 ({\color{grey}$\pm$ 0.0476}) & 0.5961 ({\color{grey}$\pm$ 0.0427}) & 0.6947 \\
       & VGG-19 & 0.8367 ({\color{grey}$\pm$ 0.0403}) & 0.7209 ({\color{grey}$\pm$ 0.0287}) & 0.7197 ({\color{grey}$\pm$ 0.0771}) & 0.6762 ({\color{grey}$\pm$ 0.0494}) & 0.6037 ({\color{grey}$\pm$ 0.0439}) & 0.6814 \\
       & ResNet-50 & 0.8492 ({\color{grey}$\pm$ 0.0412}) & 0.7651 ({\color{grey}$\pm$ 0.0249}) & 0.7561 ({\color{grey}$\pm$ 0.0688}) & 0.6814 ({\color{grey}$\pm$ 0.0482}) & 0.6542 ({\color{grey}$\pm$ 0.0374}) & 0.7183 \\
    \hline
    \multirow{7}{*} {BVQA} & VIIDEO & 0.0503 ({\color{grey}$\pm$ 0.1387}) & 0.2874 ({\color{grey}$\pm$ 0.0517}) & 0.0808 ({\color{grey}$\pm$ 0.1237}) & 0.0461 ({\color{grey}$\pm$ 0.0815}) & 0.0567 ({\color{grey}$\pm$ 0.0551}) & 0.1381 \\
      & V-BLIINDS & 0.7950 ({\color{grey}$\pm$ 0.0661}) & 0.7063 ({\color{grey}$\pm$ 0.0343}) & 0.5702 ({\color{grey}$\pm$ 0.0987}) & 0.6811 ({\color{grey}$\pm$ 0.0530}) & 0.5348 ({\color{grey}$\pm$ 0.0467}) & 0.6415 \\
      & TLVQM & 0.7799 ({\color{grey}$\pm$ 0.0494}) & 0.7588 ({\color{grey}$\pm$ 0.0260}) & 0.7849 ({\color{grey}$\pm$ 0.0650}) & 0.7878 ({\color{grey}$\pm$ 0.0341}) & 0.6568 ({\color{grey}$\pm$ 0.0418}) & 0.7323 \\
      & VIDEVAL & 0.8144 ({\color{grey}$\pm$ 0.0462}) & 0.7704 ({\color{grey}$\pm$ 0.0242}) & 0.6706 ({\color{grey}$\pm$ 0.0975}) & 0.7438 ({\color{grey}$\pm$ 0.0455}) & 0.7763 ({\color{grey}$\pm$ 0.0280}) & 0.7647 \\
      & RAPIQUE & 0.8071 ({\color{grey}$\pm$ 0.0557}) & 0.7884 ({\color{grey}$\pm$ 0.0236}) & 0.6658 ({\color{grey}$\pm$ 0.0970}) & 0.7413 ({\color{grey}$\pm$ 0.0450}) & 0.7473 ({\color{grey}$\pm$ 0.0329}) & 0.7600 \\
      & VSFA & 0.8501 ({\color{grey}$\pm$ 0.0390}) & 0.7943 ({\color{grey}$\pm$ 0.0214}) & 0.7080 ({\color{grey}$\pm$ 0.0822}) & 0.7176 ({\color{grey}$\pm$ 0.0483}) & 0.7873 ({\color{grey}$\pm$ 0.0229}) & 0.7772 \\
      & Proposed & {\bf{0.8626}} ({\color{grey}$\pm$ 0.0396}) & \bf{0.8354} ({\color{grey}\bf{$\pm$ 0.0193}}) & {\bf{0.8334}} ({\color{grey}\bf{$\pm$ 0.0622}}) & \bf{0.8414} ({\color{grey}\bf{$\pm$ 0.0276}}) & {\bf{0.8252}} ({\color{grey}\bf{$\pm$ 0.0219}}) & \bf{0.8348} \\
    \midrule
    & \bf{Criteria}  & \multicolumn{5}{c}{PLCC}\\
    \hline
    \multirow{10}{*} {BIQA} & NIQE & 0.6070 ({\color{grey}$\pm$ 0.1042}) & 0.5513 ({\color{grey}$\pm$ 0.0348}) & 0.5336 ({\color{grey}$\pm$ 0.0985}) & 0.6312 ({\color{grey}$\pm$ 0.0504}) & 0.2982 ({\color{grey}$\pm$ 0.0484}) & 0.4822 \\
       & IL-NIQE & 0.5420 ({\color{grey}$\pm$ 0.0953}) & 0.5371 ({\color{grey}$\pm$ 0.0489}) & 0.2852 ({\color{grey}$\pm$ 0.1213}) & 0.5433 ({\color{grey}$\pm$ 0.0653}) & 0.3585 ({\color{grey}$\pm$ 0.0475}) & 0.4624 \\
       & BRISQUE & 0.8049 ({\color{grey}$\pm$ 0.0616}) & 0.6513 ({\color{grey}$\pm$ 0.0405}) & 0.5986 ({\color{grey}$\pm$ 0.1031}) & 0.6242 ({\color{grey}$\pm$ 0.0611}) & 0.4073 ({\color{grey}$\pm$ 0.0612}) & 0.5713 \\
       & M3 & 0.8138 ({\color{grey}$\pm$ 0.0533}) & 0.6452 ({\color{grey}$\pm$ 0.0411}) & 0.6687 ({\color{grey}$\pm$ 0.0930}) & 0.6218 ({\color{grey}$\pm$ 0.0613}) & 0.3769 ({\color{grey}$\pm$ 0.0554}) & 0.5634 \\
       & HIGRADE & 0.7261 ({\color{grey}$\pm$ 0.0869}) & 0.7104 ({\color{grey}$\pm$ 0.0335}) & 0.6691 ({\color{grey}$\pm$ 0.0853}) & 0.6188 ({\color{grey}$\pm$ 0.0635}) & 0.7103 ({\color{grey}$\pm$ 0.0340}) & 0.6930 \\
       & FRIQUEE & 0.8415 ({\color{grey}$\pm$ 0.0433}) & 0.7354 ({\color{grey}$\pm$ 0.0265}) & 0.7481 ({\color{grey}$\pm$ 0.0842}) & 0.6914 ({\color{grey}$\pm$ 0.0595}) & 0.7505 ({\color{grey}$\pm$ 0.0283}) & 0.7410 \\
       & CORNIA & 0.6631 ({\color{grey}$\pm$ 0.0765}) & 0.7356 ({\color{grey}$\pm$ 0.0257}) & 0.5203 ({\color{grey}$\pm$ 0.0984}) & 0.7239 ({\color{grey}$\pm$ 0.0379}) & 0.5851 ({\color{grey}$\pm$ 0.0396}) & 0.6642 \\
       & HOSA & 0.8673 ({\color{grey}\bf{$\pm$ 0.0312}}) & 0.7580 ({\color{grey}$\pm$ 0.0244}) & 0.7451 ({\color{grey}$\pm$ 0.0731}) & 0.7242 ({\color{grey}$\pm$ 0.0376}) & 0.6037 ({\color{grey}$\pm$ 0.0417}) & 0.7066 \\
       & VGG-19 & 0.8505 ({\color{grey}$\pm$ 0.0406}) & 0.7385 ({\color{grey}$\pm$ 0.0263}) & 0.7594 ({\color{grey}$\pm$ 0.0705}) & 0.7281 ({\color{grey}$\pm$ 0.0446}) & 0.6074 ({\color{grey}$\pm$ 0.0441}) & 0.7013 \\
       & ResNet-50 & 0.8652 ({\color{grey}$\pm$ 0.0392}) & 0.7781 ({\color{grey}$\pm$ 0.0232}) & 0.7957 ({\color{grey}$\pm$ 0.0601}) & 0.7381 ({\color{grey}$\pm$ 0.0395}) & 0.6485 ({\color{grey}$\pm$ 0.0413}) & 0.7344 \\
       \hline
    \multirow{7}{*} {BVQA} & VIIDEO & 0.2479 ({\color{grey}$\pm$ 0.1035}) & 0.3083 ({\color{grey}$\pm$ 0.0480}) & 0.2301 ({\color{grey}$\pm$ 0.0980}) & 0.2100 ({\color{grey}$\pm$ 0.0720}) & 0.1497 ({\color{grey}$\pm$ 0.0544}) & 0.2284 \\
       & V-BLIINDS & 0.8067 ({\color{grey}$\pm$ 0.0761}) & 0.7011 ({\color{grey}$\pm$ 0.0342}) & 0.6269 ({\color{grey}$\pm$ 0.0881}) & 0.6997 ({\color{grey}$\pm$ 0.0499}) & 0.5409 ({\color{grey}$\pm$ 0.0461}) & 0.6493 \\
       & TLVQM & 0.7904 ({\color{grey}$\pm$ 0.0499}) & 0.7598 ({\color{grey}$\pm$ 0.0254}) & 0.8152 ({\color{grey}$\pm$ 0.0655}) & 0.7942 ({\color{grey}$\pm$ 0.0339}) & 0.6470 ({\color{grey}$\pm$ 0.0406}) & 0.7331 \\
       & VIDEVAL & 0.8320 ({\color{grey}$\pm$ 0.0538}) & 0.7709 ({\color{grey}$\pm$ 0.0273}) & 0.7054 ({\color{grey}$\pm$ 0.1035}) & 0.7476 ({\color{grey}$\pm$ 0.0445}) & 0.7715 ({\color{grey}$\pm$ 0.0290}) & 0.7673 \\
       & RAPIQUE & 0.8232 ({\color{grey}$\pm$ 0.0529}) & 0.8051 ({\color{grey}$\pm$ 0.0217}) & 0.6913 ({\color{grey}$\pm$ 0.0916}) & 0.7618 ({\color{grey}$\pm$ 0.0408}) & 0.7569 ({\color{grey}$\pm$ 0.0310}) & 0.7755 \\
       & VSFA & 0.8690 ({\color{grey}$\pm$ 0.0379}) & 0.7985 ({\color{grey}$\pm$ 0.0207}) & 0.7741 ({\color{grey}$\pm$ 0.0724}) & 0.7707 ({\color{grey}$\pm$ 0.0379}) & 0.7888 ({\color{grey}\bf{$\pm$ 0.0225}}) & 0.7938 \\
       & Proposed & {\bf{0.8826}} ({\color{grey}$\pm$ 0.0372}) & \bf{0.8339} ({\color{grey}\bf{$\pm$ 0.0178}}) & {\bf{0.8371}} ({\color{grey}\bf{$\pm$ 0.0517}}) & \bf{0.8394} ({\color{grey}\bf{$\pm$ 0.0284}}) & {\bf{0.8178}} ({\color{grey}$\pm$ 0.0260}) & \bf{0.8330} \\
     \bottomrule
   \end{tabular}
   \label{Table:SingleDatabasePerform_onCKLNY}
\end{table*}

\begin{table}[!htbp]
   \small
	\centering
	\caption{SRCC and PLCC results on LSVQ~\cite{ying2021patch} under the individual database training setting. The database size is shown in the bracket}
	\scalebox{0.93}{
	\setlength{\tabcolsep}{1.0mm}{
		\begin{tabular}{l|l|cc|cc|cc}
			\toprule
			\multirow{2}{*} {} & \bf{Database} & \multicolumn{2}{c|} {Test(7.4k)} & \multicolumn{2}{c|} {Test-1080p(3.5k)} & \multicolumn{2}{c} {W.A.(10.9k)} \\
            \hline
			& \bf{Criteria} & SRCC & PLCC & SRCC & PLCC & SRCC & PLCC \\
			\hline
			BIQA & BRISQUE     & 0.579  & 0.576  & 0.497  & 0.531 & 0.5527 & 0.5616 \\
            \hline
            \multirow{6}{*} {BVQA} & TLVQM       & 0.772  & 0.774  & 0.589  & 0.616 & 0.7132 & 0.7233 \\
            & VIDEVAL     & 0.794  & 0.783  & 0.545  & 0.554 & 0.7140 & 0.7095 \\
            & VSFA        & 0.801  & 0.796  & 0.675  & 0.704 & 0.7605 & 0.7665 \\
            & PVQ(w/o)    & 0.814 & 0.816 & 0.686 & 0.708 & 0.7729 & 0.7813 \\
            & PVQ(w)      & 0.827 & 0.828 & 0.711 & 0.739 & 0.7898 & 0.7994 \\
            & Proposed    & \bf{0.852} & \bf{0.854} & \bf{0.772} & \bf{0.788} & \bf{0.8261} & \bf{0.8324} \\
            \bottomrule
	\end{tabular}
    }}
	\label{Table:SingleDatabasePerform_onQ}
\end{table}

We conduct experiments on six in-the-wild VQA databases, \textit{i.e.}, CVD2014~\cite{nuutinen2016cvd2014}, KoNViD-1k~\cite{hosu2017konstanz}, LIVE-Qualcomm~\cite{ghadiyaram2018capture}, LIVE-VQC~\cite{sinno2019large}, YouTube-UGC~\cite{wang2019youtube}, and LSVQ~\cite{ying2021patch}. The main information of these databases are summarized in Table~\ref{Table:VQAdatabase}. It is clear that they differ in content, resolution, time duration, and annotation scale, \textit{etc}.

\subsubsection{Competing Methods}\label{subsec:ComparedAlgs}
We compare the proposed method against both adapted BIQA and BVQA models. Note that we evolve BIQA algorithms into the baselines for VQA by extracting frame-level features, followed by temporal average pooling to obtain the video-level features. The representative BIQA models are NIQE~\cite{mittal2013making}, IL-NIQE~\cite{zhang2015feature}, BRISQUE~\cite{mittal2012no}, M3~\cite{xue2014blind}, HIGRADE~\cite{kundu2017no}, FRIQUEE~\cite{ghadiyaram2017perceptual}, CORNIA~\cite{ye2012unsupervised}, HOSA~\cite{xu2016blind}, and pre-trained DNN models, VGG-19~\cite{simonyan2014very}, ResNet-50~\cite{he2016deep}. The compared BVQA methods can be roughly classified into three groups: 1), an opinion-unaware model, \textit{i.e.}, VIIDEO~\cite{mittal2016completely}; 2), three hand-crafted models: V-BLIINDS~\cite{saad2014blind}, TLVQM~\cite{korhonen2019two}, and VIDEVAL~\cite{tu2021ugc}; and 3), four DNN methods: VSFA~\cite{li2019quality}, MDTVSFA~\cite{li2021unified}, RAPIQUE~\cite{tu2021rapique}, and PVQ~\cite{ying2021patch}.

\subsubsection{Performance Criteria}\label{subsec:Criteria}
We use two criteria to benchmark all methods. Specifically, the SRCC is used to measure the prediction monotonicity. The PLCC is adopted to evaluate the prediction accuracy. Before calculating PLCC, a nonlinear logistic mapping is applied as suggested in~\cite{video2000final}. Here, we employ a four-parameter logistic function. Except for LSVQ, we split each database into 60\% for training, 20\% for validation, and 20\% for testing with no overlap of video contents. Apart from three training-free methods NIQE, IL-NIQE, and VIIDEO, we re-train and validate the remaining methods on the same training/validation/testing splits. For the main experimental results summarized in Table~\ref{Table:SingleDatabasePerform_onCKLNY}, \textit{i.e.}, evaluation under the individual database training setting, we randomly repeat this procedure 100 times to prevent performance bias. The other experiments are conducted using the first ten seeds of the above 100 repetitions. Finally, the median results are recorded for comparison. As for LSVQ, we follow~\cite{ying2021patch} to train the model on a training set and evaluate it on two test subsets. We also provide database-size weighted average results (abbreviated as W.A.) to give insight into overall performance across different databases.

\subsubsection{Implementation Details}\label{subsec:Impldetail}
To train the frame-level feature extractor, we use the pre-trained ResNet-50 on ImageNet~\cite{deng2009imagenet} to initialize the backbone network. Following~\cite{zhang2021uncertainty}, we set the parameter $\eta$ to 0.025 and $\nu$ to 1. We minimize $\ell_{s}$ using 250,000 image pairs randomly sampled from the IQA databases with the resolution of 384$\times$384$\times$3. We train the model for $12$ epochs with a learning rate decay factor of 10 for every 3 epochs from an initialization of $10^{-4}$. For motion feature extraction, we apply the SlowFast pre-trained on Kinetics-400~\cite{kay2017kinetics} as stated in Section~\ref{subsection:FeaturesFusion}. During fine-tuning on the target VQA databases, the weights of the above two sub-networks are frozen. We set the hidden size of the GRU to 32. The duration $\tau$ and equilibrium factor $\beta$ in the hysteresis-based temporal pooling are set to 12 and 0.5, respectively. All the learnable parameters are optimized using Adam~\cite{Kingma2014adam} with a mini-batch of 32, an iteration of 40 epochs, and an initial learning rate of $5\times10^{-4}$ which decays with a ratio of 0.2 for every two epochs. And the balanced factor $\lambda$ in the loss function is set to 1. The proposed method is implemented using PyTorch, and the source code is available at \url{https://github.com/zwx8981/BVQA-2021}.

\subsection{Performance on Individual Databases}\label{subsec:IndividualExp}
We compare the performance on each single database in Table~\ref{Table:SingleDatabasePerform_onCKLNY} and Table~\ref{Table:SingleDatabasePerform_onQ}, from which we have several interesting observations. First, adapted BIQA methods attain meaningful performance on VQA databases, which validates their promises for VQA to some extent. VGG-19 and ResNet-50 demonstrate promising results, which suggest the advantage of the data-driven features learned from large-scale databases over knowledge-based features. Second, by incorporating temporal information, BVQA methods generally perform competitively against adapted BIQA models, which suggests that temporal modeling is valuable for BVQA. One exception is VIIDEO, which is calibrated to handle synthetic distortions. VSFA relies on frame-level features of pre-trained DNNs and models the temporal information using a RNN, yielding competitive performance on all databases. By incorporating local and global features in a unified framework and using a 3D network to capture temporal distortions, PVQ achieves significant performance improvement on LSVQ. Third, the proposed method learns more perceptually meaningful spatial features via a quality-aware pre-training process, transfers the motion-related knowledge from a 3D-CNN optimized on an action recognition task, and fine-tunes the entire model using a mixed list-wise ranking loss function. In general, our method achieves superior performance on all in-the-wild VQA databases. Typically, the proposed method outperforms VSFA on LIVE-VQC by a large margin (+$12.38\%$ SRCC). It also presents remarkable performance on larger databases (KoNViD-1k, YouTube-UGC, and LSVQ), indicating its promising representation learning capacity on large-scale applications.

\begin{table}[!htbp]
	\centering
	\caption{Median SRCC and PLCC results under the mixed databases training setting}
	\scalebox{0.935}{
	\setlength{\tabcolsep}{2.0mm}{
		\begin{tabular}{l|c|c|c|c}
			\toprule
			\bf{Database} & \bf{Criteria} & \multicolumn{1}{c|} {MDTVSFA} & \multicolumn{1}{c|} {Proposed-LS} & \multicolumn{1}{c} {Proposed} \\

			\hline
			\multirow{2}{*} {CVD2014}       & SRCC & 0.8326 & 0.8406 & \bf{0.8737} \\
			                                & PLCC & 0.8347 & 0.8519 & \bf{0.8793} \\
            \hline
			\multirow{2}{*} {KoNViD-1k}     & SRCC & 0.7816 & 0.7976 & \bf{0.8205} \\
			                                & PLCC & 0.7786 & 0.7896 & \bf{0.8142} \\
            \hline
			\multirow{2}{*} {LIVE-Qualcomm} & SRCC & 0.8136 & 0.8041 & \bf{0.8387} \\
			                                & PLCC & 0.8291 & 0.8253 & \bf{0.8617} \\
            \hline
			\multirow{2}{*} {LIVE-VQC}      & SRCC & 0.7277 & 0.7889 & \bf{0.8162} \\
			                                & PLCC & 0.7784 & 0.7806 & \bf{0.8445} \\
            \hline
			\multirow{2}{*} {W.A.} & SRCC & 0.7779 & 0.7993 & \bf{0.8307} \\
			                                   & PLCC & 0.7860 & 0.7974 & \bf{0.8260} \\

            \bottomrule
	\end{tabular}}}
	\label{Table:DatabaseMixedTest_onCKLN}
\end{table}

\subsection{Performance on Mixed Databases}\label{subsec:MixedPerform}
\subsubsection{Database-level Mixed Test}\label{subsubsec:DatabaseMixedTest}
In a more practical experimental setting, a BVQA model is expected to perform well across different data distributions. Similar to MDTVSFA~\cite{li2021unified}, we adopt a database combination strategy to train the proposed method on mixed databases. We compare with MDTVSFA by mixing the four databases including CVD2014, KoNViD-1k, LIVE-Qualcomm, and LIVE-VQC for training. The results are summarized in Table~\ref{Table:DatabaseMixedTest_onCKLN}. We also apply the naive linear re-scaling (LS)~\cite{korhonen2019two} to integrate the subjective quality scores and use the L1 loss function for regression, termed as Proposed-LS. We observe that our BVQA metric consistently outperforms MDTVSFA. The mixed list-wise ranking method is superior to LS in terms of the overall performance on all databases, indicating favorable perceptual alignment across multiple databases with different scales of annotations. Taking a comparison of Table~\ref{Table:DatabaseMixedTest_onCKLN} and Table~\ref{Table:SingleDatabasePerform_onCKLNY}, we observe performance improvements on CVD2014 and LIVE-Qualcomm yet slight drops on KoNViD-1k and LIVE-VQC. This shows the mixed databases training strategy may help alleviating the over-fitting phenomenon on small databases while bringing acceptable disturbance to the larger ones.

\begin{table*}[!htbp]
	\centering
	\caption{Median SRCC and PLCC results on different categories of mixed subsets, including resolution types of 1080p, 720p, and $\le$480p,  content types of screen, animation, and gaming, and quality types of low quality and high quality from KoNViD-1k~\cite{hosu2017konstanz}, LIVE-VQC~\cite{sinno2019large}, and YouTube-UGC~\cite{wang2019youtube}}
	\setlength{\tabcolsep}{0.7mm}{
		\begin{tabular}{l|cc|cc|cc|cc|cc|cc|cc|cc|cc}
			\toprule
            {\bf Category} & \multicolumn{6}{c|} {Resolution} & \multicolumn{6}{c|} {Content} & \multicolumn{4}{c|} {Quality} & \multicolumn{2}{c} {} \\
            \hline
			{\bf Subset} & \multicolumn{2}{c|} {1080p} & \multicolumn{2}{c|} {720p} & \multicolumn{2}{c|} {$\le$480p} & \multicolumn{2}{c|} {Screen} & \multicolumn{2}{c|} {Animation} & \multicolumn{2}{c|} {Gaming} & \multicolumn{2}{c|} {Low Quality} & \multicolumn{2}{c|} {High Quality} & \multicolumn{2}{c} {W.A.} \\
			\cline{1-3} \cline{4-5} \cline{6-7} \cline{8-9} \cline{10-11} \cline{12-13} \cline{14-15} \cline{16-17} \cline{18-19}
			{\bf Criteria} & SRCC & PLCC & SRCC & PLCC & SRCC & PLCC & SRCC & PLCC & SRCC & PLCC & SRCC & PLCC & SRCC & PLCC & SRCC & PLCC & SRCC & PLCC \\
			\hline
            BRISQUE   & 0.4447 & 0.4645 & 0.5972 & 0.5950 & 0.5092 & 0.5088 & 0.1770 & 0.3937 & 0.0945 & 0.4862 & 0.3200 & 0.3605 & 0.4347 & 0.4617 & 0.2986 & 0.3110 & 0.4067 & 0.4329\\
            M3        & 0.4541 & 0.4958 & 0.5129 & 0.4933 & 0.4989 & 0.5180 & 0.2167 & 0.2387 & 0.2813 & 0.5212 & 0.2505 & 0.3262 & 0.3992 & 0.4840 & 0.2384 & 0.2328 & 0.3691 & 0.4033 \\
            HIGRADE   & 0.4574 & 0.5316 & 0.5447 & 0.5478 & 0.6147 & 0.6282 & 0.4603 & 0.5555 & 0.2703 & 0.4348 & 0.5692 & 0.6233 & 0.5463 & 0.5707 & 0.4659 & 0.4792 & 0.5178 & 0.5442 \\
            FRIQUEE   & 0.5513 & 0.6039 & 0.6130 & 0.6118 & 0.6735 & 0.6966 & 0.4779 & 0.5612 & 0.1846 & 0.4929 & 0.6527 & 0.6944 & 0.5319 & 0.5926 & 0.5004 & 0.5360 & 0.5490 & 0.5932\\
            CORNIA    & 0.6140 & \bf{0.7063} & 0.6176 & 0.6190 & 0.6667 & 0.7080 & 0.3304 & 0.4187 & 0.2461 & 0.4674 & 0.5241 & 0.6060 & 0.4867 & 0.5454 & 0.3619 & 0.3712 & 0.4911 & 0.5330\\
            HOSA      & 0.5747 & 0.6423 & 0.6578 & 0.6489 & 0.7158 & 0.7302 & 0.3396 & 0.5127 & 0.1329 & 0.4591 & 0.5611 & 0.6057 & 0.5411 & 0.5874 & 0.4296 & 0.4402 & 0.5351 & 0.5696\\
            VGG-19    & 0.6240 & 0.6546 & 0.6407 & 0.6370 & 0.7076 & 0.7138 & 0.4025 & 0.4689 & 0.1340 & 0.4182 & 0.5436 & 0.5841 & 0.5528 & 0.5969 & 0.4332 & 0.4375 & 0.5419 & 0.5667\\
            ResNet-50 & 0.6373 & 0.6713 & 0.6548 & 0.6781 & 0.7591 & 0.7745 & 0.4213 & 0.5220 & 0.2549 & 0.4270 & 0.5639 & 0.5945 & 0.5944 & 0.6476 & 0.4621 & 0.4701 & 0.5751 & 0.6073 \\
            \hline
            V-BLIINDS & 0.4048 & 0.5063 & 0.5642 & 0.5758 & 0.5912 & 0.6096 & 0.1730 & 0.3036 & -0.1560 & 0.4471 & 0.4138 & 0.5447 & 0.5033 & 0.5095 & 0.3503 & 0.3545 & 0.4457 & 0.4778 \\
            TLVQM     & 0.5530 & 0.6329 & 0.6464 & 0.6467 & 0.6113 & 0.6332 & 0.3266 & 0.4690 & 0.1274 & 0.3926 & 0.6014 & 0.6213 & 0.5026 & 0.5497 & 0.5064 & 0.5130 & 0.5314 & 0.5653\\
            VIDEVAL   & 0.5536 & 0.6016 & 0.6312 & 0.6381 & 0.6106 & 0.6667 & 0.5307 & 0.6469 & 0.2109 & 0.4480 & 0.6971 & 0.7105 & 0.5816 & 0.6194 & 0.5558 & 0.5830 & 0.5785 & 0.6168 \\
            RAPIQUE   & 0.5262 & 0.6230 & 0.5911 & 0.6362 & 0.6820 & 0.6927 & 0.4392 & 0.4514 & 0.3120 & 0.4882 & 0.5471 & 0.6643 & 0.6349 & 0.6828 & 0.5250 & 0.5391 & 0.5807 & 0.6210 \\
            VSFA      & 0.6512 & 0.6697 & 0.6677 & 0.6650 & 0.7084 & 0.7277 & 0.4700 & 0.5493 & -0.0021 & 0.4355 & 0.6774 & 0.7501 & 0.6169 & 0.6741 & 0.5286 & 0.5313 & 0.5999 & 0.6327 \\
            Proposed  & \bf{0.6759} & 0.6902 & \bf{0.7431} & \bf{0.7333} & \bf{0.7819} & \bf{0.7972} & \bf{0.5726} & \bf{0.6845} & \bf{0.5604} & \bf{0.6635} & \bf{0.7073} & \bf{0.7655} & \bf{0.6863} & \bf{0.7118} & \bf{0.6524} & \bf{0.6694} & \bf{0.6896} & \bf{0.7110} \\
            \bottomrule
	\end{tabular}}
	\label{Table:CategoryMixedTest}
\end{table*}

\subsubsection{Category-level Mixed Test}\label{subsubsec:CategoryMixedTest}
We follow the three categorical evaluation methodologies in~\cite{tu2021ugc} to give insights into different aspects. To this end, we combine KoNViD-1k, LIVE-VQC, and YouTube-UGC calibrated via INLSA~\cite{pinson2003objective} for experiments, termed as Combined$_{U}$.

\paragraph{Resolution Subsets}\label{subsubsec:ResolutionCategoryMixedTest}
According to resolution, the Combined$_{U}$ can be formed into three sets, \textit{i.e.}, 402 1080p-videos, 564 720p-videos, and 607 videos with resolution no more than 480p. We list the results in the ``Resolution'' column of Table~\ref{Table:CategoryMixedTest}.

\paragraph{Content Subsets}\label{subsubsec:ContentCategoryMixedTest}
There is plenty of researches concentrating on different content-based scenarios. It is also interesting to observe the behaviors of compared models on different contents. To this end, we conduct experiments using three typical contents, \textit{i.e.}, 134 Screens, 70 Animations, and 180 Gamings. We report the performance in the ``Content'' column of Table~\ref{Table:CategoryMixedTest}.

\paragraph{Quality Subsets}\label{subsubsec:QualityCategoryMixedTest}
The partition based on low and high quality is a valuable way to analyze the defects and success of the proposed model, which has been adopted as an evaluation methodology in both IQA~\cite{yu2019predicting} and VQA~\cite{tu2021ugc}. We use the threshold of 3.5537~\cite{tu2021ugc} to partition Combined$_{U}$ into 1,469 low quality and 1,458 high quality materials, and tabulate the comparisons in the ``Quality'' column of Table~\ref{Table:CategoryMixedTest}.

The results in Table~\ref{Table:CategoryMixedTest} manifest the proposed method is effective and robust across different resolutions, contents, and quality levels. From the ``Resolution'' column of Table~\ref{Table:CategoryMixedTest}, we can see that learning-based features are more powerful than hand-crafted features. However, the ``Content'' and ``Quality'' columns of Table~\ref{Table:CategoryMixedTest} demonstrate that pre-trained DNNs on an image classification task struggle to handle scenarios with different contents and quality levels. This further verifies the effectiveness of the proposed quality-aware pre-training and motion perception schemes.

\begin{table*}[htbp!]
  \centering
  \caption{SRCC and PLCC results of the cross-database evaluation on CVD2014~\cite{nuutinen2016cvd2014}, KoNViD-1k~\cite{hosu2017konstanz}, LIVE-Qualcomm~\cite{ghadiyaram2018capture}, LIVE-VQC~\cite{sinno2019large}, and YouTube-UGC~\cite{wang2019youtube}}\label{Tab:CrossPerform_T2}
  \setlength{\tabcolsep}{0.6mm}{
  \begin{tabular}{l|cc|cc|cc|cc|cc|cc|cc|cc|cc}
      \toprule
        \bf{Training} & \multicolumn{2}{c|} {CVD2014} & \multicolumn{2}{c|} {LIVE-Qualcomm} & \multicolumn{4}{c|} {KoNViD-1k} & \multicolumn{4}{c|} {LIVE-VQC} & \multicolumn{4}{c|} {YouTube-UGC} & \multicolumn{2}{c} {} \\
     \hline
        \bf{Testing}  & \multicolumn{2}{c|} {LIVE-Qualcomm} & \multicolumn{2}{c|} {CVD2014} & \multicolumn{2}{c|} {LIVE-VQC} & \multicolumn{2}{c|} {YouTube-UGC} & \multicolumn{2}{c|} {KoNViD-1k} & \multicolumn{2}{c|} {YouTube-UGC} & \multicolumn{2}{c|} {KoNViD-1k} & \multicolumn{2}{c|} {LIVE-VQC} & \multicolumn{2}{c} {W.A.} \\
     \hline
        \bf{Criteria} & SRCC & PLCC & SRCC & PLCC & SRCC & PLCC & SRCC & PLCC & SRCC & PLCC & SRCC & PLCC & SRCC & PLCC & SRCC & PLCC & SRCC & PLCC \\
     \hline
        BRISUQE  & 0.4732 & 0.5141 & 0.2550 & 0.2567 & 0.6160 & 0.6492 & 0.5682 & 0.6052 & 0.5702 & 0.5726 & 0.1519 & 0.1666 & 0.5175 & 0.5065 & 0.4573 & 0.4841 & 0.4628 & 0.4775 \\
        TLVQM    & 0.4993 & 0.5238 & 0.3920 & 0.4255 & 0.5724 & 0.6293 & 0.7206 & 0.7530 & 0.6398 & 0.6310 & 0.2181 & 0.2501 & 0.5558 & 0.5783 & 0.4883 & 0.5463 & 0.5278 & 0.5548 \\
        VIDEVAL  & 0.4902 & 0.5016 & \bf{0.5784} & \bf{0.5811} & 0.5916 & 0.6289 & 0.7022 & 0.7153 & 0.6565 & 0.6533 & 0.2390 & 0.2612 & 0.6839 & 0.6863 & 0.4399 & 0.4839 & 0.5597 & 0.5740\\
        VSFA     & 0.4129 & 0.4840 & 0.4306 & 0.4666 & 0.5934 & 0.6061 & 0.7175 & 0.7597 & 0.6949 & 0.7109 & 0.4221 & 0.4525 & 0.7167 & 0.7111 & 0.6257 & 0.6597 & 0.6187 & 0.6418 \\
        Proposed & \bf{0.6696} & \bf{0.7007} & 0.4701 & 0.4807 & \bf{0.6949} & \bf{0.7120} & \bf{0.7799} & \bf{0.7803} & \bf{0.7382} & \bf{0.7210} & \bf{0.6025} & \bf{0.6029} & \bf{0.7847} & \bf{0.7818} & \bf{0.6887} & \bf{0.7274} & \bf{0.7092} & \bf{0.7121} \\
     \bottomrule
   \end{tabular}}
   \label{Table:CrossEvaluation_onCKLNY}
\end{table*}

\begin{table}[htbp!]
  \centering
  \caption{SRCC and PLCC results of the cross-database evaluation where the model is trained on LSVQ~\cite{ying2021patch}, and then tested on KoNViD-1k~\cite{hosu2017konstanz} and LIVE-VQC~\cite{sinno2019large}}\label{Tab:CrossPerform_T3}
  \begin{tabular}{l|cc|cc|cc}
      \toprule
        \bf{Testing} & \multicolumn{2}{c|} {KoNViD-1k} & \multicolumn{2}{c|} {LIVE-VQC} & \multicolumn{2}{c} {W.A.} \\
     \hline
        \bf{Criteria} & SRCC & PLCC & SRCC & PLCC & SRCC & PLCC \\
     \hline
        BRISQUE  & 0.646 & 0.647 & 0.524 & 0.536 & 0.6060 & 0.6106 \\
        TLVQM    & 0.732 & 0.724 & 0.670 & 0.691 & 0.7117 & 0.7132 \\
        VIDEVAL  & 0.751 & 0.741 & 0.630 & 0.640 & 0.7113 & 0.7079 \\
        VSFA     & 0.784 & 0.794 & 0.734 & 0.772 & 0.7676 & 0.7868 \\
        PVQ(w/o) & 0.782 & 0.781 & 0.747 & 0.776 & 0.7705 & 0.7794 \\
        PVQ(w)   & 0.791 & 0.795 & 0.770 & 0.807 & 0.7841 & 0.7989 \\
        Proposed & \bf{0.839} & \bf{0.830} & \bf{0.816} & \bf{0.824} & \bf{0.8315} & \bf{0.8280} \\
     \bottomrule
   \end{tabular}
   \label{Table:CrossEvaluation_onQ}
\end{table}

\begin{figure}[htbp!]
    \centering
    \captionsetup{justification=centering}
    \subfloat[Three representative frames of Video A in YouTube-UGC]{\includegraphics[width=1\columnwidth]{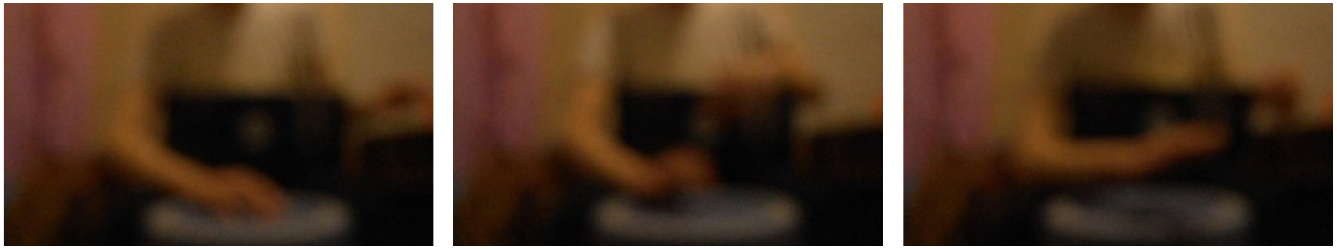}}
    \vspace{2pt}
    \subfloat[Three representative frames of Video B in YouTube-UGC]{\includegraphics[width=1\columnwidth]{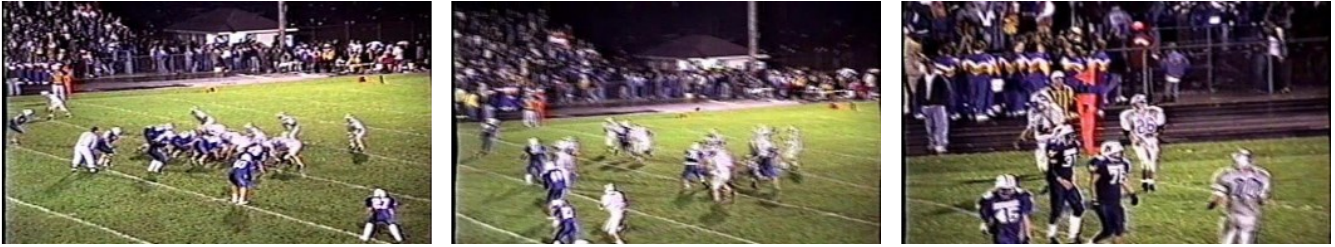}}
    \vspace{2pt}
    \subfloat[Three representative frames of Video C in YouTube-UGC]{\includegraphics[width=1\columnwidth]{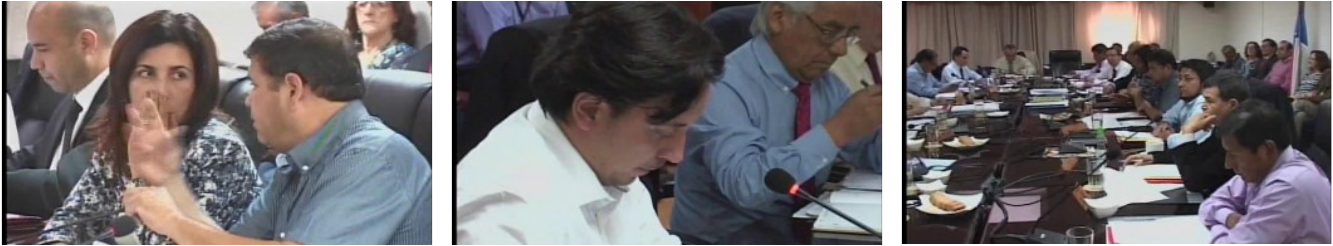}}
    \vspace{2pt}
    \subfloat[Three representative frames of Video D in YouTube-UGC]{\includegraphics[width=1\columnwidth]{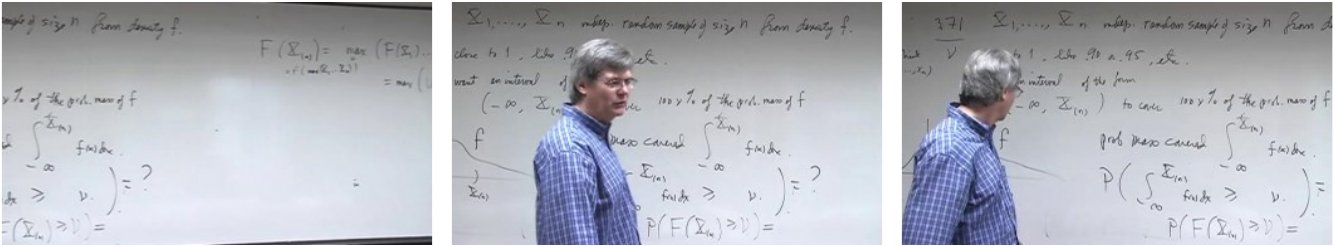}}
    \vspace{2pt}
    \subfloat[Three representative frames of Video E in YouTube-UGC]{\includegraphics[width=1\columnwidth]{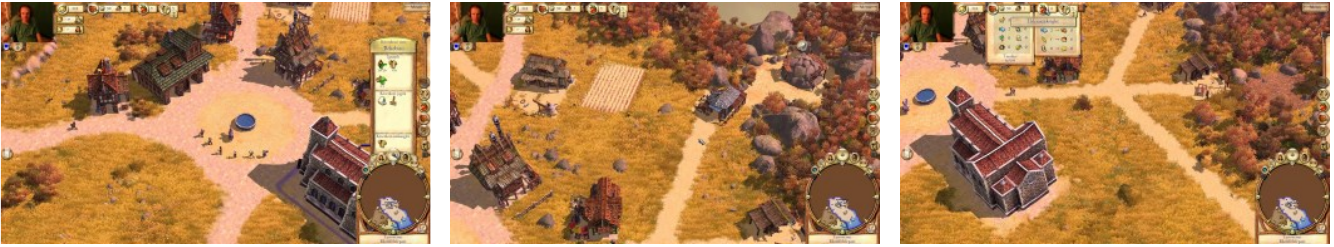}}
    \vspace{2pt}
    \subfloat[Three representative frames of Video F in YouTube-UGC]{\includegraphics[width=1\columnwidth]{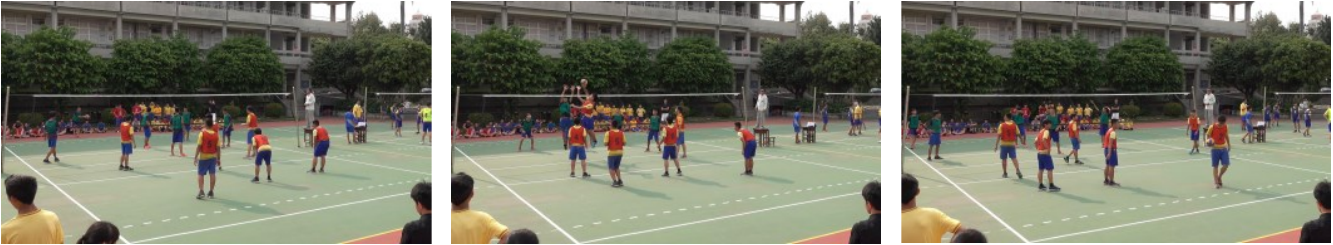}}
    \vspace{2pt}
    \caption{Successful cases sampled from the YouTube-UGC~\cite{wang2019youtube} test set. The MOSs of A to F are 1.462, 1.706, 2.983, 3.033, 4.399, and 4.526, respectively. The predictions of our model are 1.760, 2.527, 3.185, 3.277, 4.311, and 4.453, respectively. Both groups of scores conform to the ranking of A{\textless}B{\textless}C{\textless}D{\textless}E{\textless}F.}\label{Figure:SucessCase}
\end{figure}

\begin{figure}[htbp!]
    \centering
    \captionsetup{justification=centering}
    \subfloat[Three representative frames of Video A in YouTube-UGC]{\includegraphics[width=1\columnwidth]{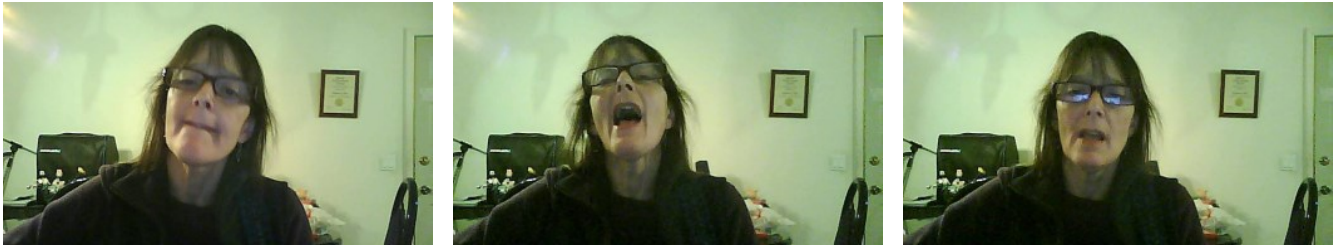}}
    \vspace{2pt}
    \subfloat[Three representative frames of Video B in YouTube-UGC]{\includegraphics[width=1\columnwidth]{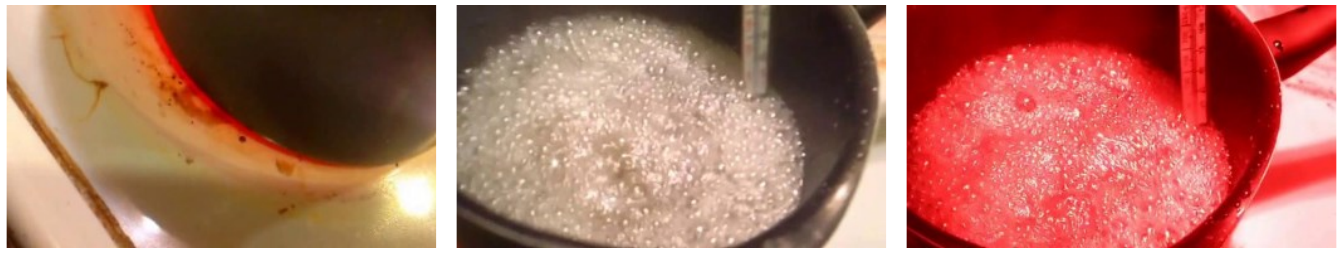}}
    \vspace{2pt}
    \subfloat[Three representative frames of Video C in KoNViD-1k]{\includegraphics[width=1\columnwidth]{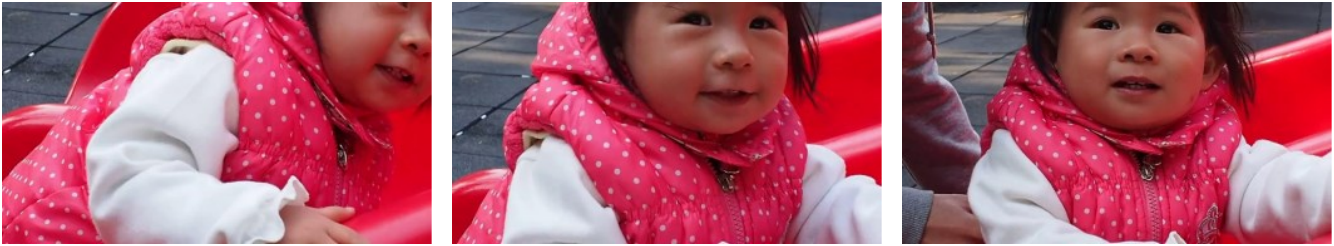}}
    \vspace{2pt}
    \subfloat[Three representative frames of Video D in KoNViD-1k]{\includegraphics[width=1\columnwidth]{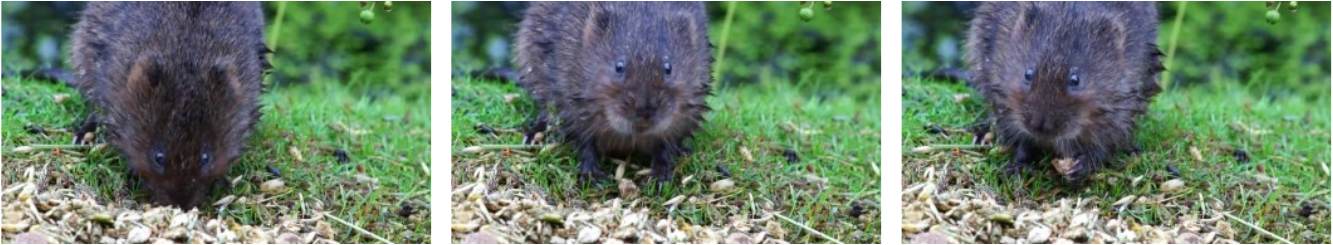}}
    \vspace{2pt}
    \caption{Failure cases sampled from the YouTube-UGC~\cite{wang2019youtube} and KoNViD-1k~\cite{hosu2017konstanz} test sets. The MOSs of A and B are both 3.056 while the predictions are 3.464 and 3.187, respectively. The MOSs of C and D are 3.88 and 4.40 while the predictions are 4.017 and 3.938, respectively.}\label{Figure:FailureCase}
\end{figure}

\subsection{Cross-database Evaluation}\label{subsec:CrossEvaluation}
A BVQA model is expected to generalize well to unseen distortion scenarios. In this regard, we conduct cross-database evaluation by training BVQA models on one database and testing them on the other databases. We report the results in Table~\ref{Table:CrossEvaluation_onCKLNY} and Table~\ref{Table:CrossEvaluation_onQ}, from which We observe that the proposed method performs well to unseen databases. Specifically, on small and medium scale databases, the improvement of average performance is over $9\%$ (SRCC). For the largest database LSVQ, our method achieves a more than $4\%$ (SRCC) improvement. These performance gains demonstrate the favorable generalizability of the proposed method, which we believe is mainly due to the effectiveness of the proposed knowledge transferring strategy.

\subsection{Qualitative Results}\label{subsec:QualitativeAnalysis}

In this subsection, we present several successful and failure samples in Fig.~\ref{Figure:SucessCase} and Fig.~\ref{Figure:FailureCase}, respectively. As shown in Fig.~\ref{Figure:SucessCase}, our method can distinguish quality levels even with small differences of MOSs. Then, we visualize several failure cases in Fig.~\ref{Figure:FailureCase}. Among them, for the two comparisons of Video A against Video B and Video C against Video D, our method still makes reasonable quality predictions although their relative rankings are not consistent with the MOSs. Note that both cases are of relatively high perceptual quality, which poses a great challenge to BVQA models for quality discrimination.

\begin{table*}[!htbp]
	\centering
	\caption{Ablation study of different model designings}
	\setlength{\tabcolsep}{1.5mm}{
		\begin{tabular}{l|l|cc|cc|cc|cc|cc|cc}
			\toprule
			& \bf{Database} & \multicolumn{2}{c|} {CVD2014} & \multicolumn{2}{c|} {KoNViD-1k} & \multicolumn{2}{c|} {LIVE-Qualcomm} & \multicolumn{2}{c|}{LIVE-VQC} & \multicolumn{2}{c|} {YouTube-UGC} & \multicolumn{2}{c} {W.A.} \\
			\cline{2-14}
			& \bf{Criteria} & SRCC & PLCC & SRCC & PLCC & SRCC & PLCC & SRCC & PLCC & SRCC & PLCC & SRCC & PLCC \\
			\hline
			\multirow{5}{*} {Feature} & baseline                           & \textbf{0.8725} & \textbf{0.8860} & 0.7735 & 0.7891 & 0.7533 & 0.7968 & 0.7107 & 0.7517 & 0.7864 & 0.7861 & 0.7726 & 0.7888 \\
            & S-Feature                          & 0.8516 & 0.8674 & 0.8272 & 0.8290 & 0.8184 & 0.8311 & 0.7752 & 0.7877 & 0.8143 & 0.8127 & 0.8149 & 0.8191 \\
            & S-Feature+SVR                      & 0.7583 & 0.7890 & 0.8182 & 0.8175 & 0.7400 & 0.7936 & 0.7459 & 0.7781 & 0.8013 & 0.8033 & 0.7909 & 0.8024 \\
			& M-Feature                          & 0.7741 & 0.7724 & 0.6649 & 0.6631 & 0.6389 & 0.6465 & 0.7245 & 0.7217 & 0.6624 & 0.6566 & 0.6804 & 0.6776 \\
            & S+M-Feature (Proposed)                       & 0.8675 & 0.8717 & \textbf{0.8362} & \textbf{0.8335} & \textbf{0.8361} & \textbf{0.8389} & \textbf{0.8412} & \textbf{0.8415} & \textbf{0.8233} & \textbf{0.8228} & \textbf{0.8349} & \textbf{0.8342} \\
           \cline{1-14}
            \multirow{3}{*} {Interaction} & S+M+SF$_S$-Feature & 0.8613 & 0.8823 & 0.8310 & 0.8326 & 0.8189 & 0.8244 & 0.8355 & 0.8408 & 0.8066 & 0.8076 & 0.8249 & 0.8285 \\
            & S+ST$_{3D}$-Feature & \textbf{0.8790} & \textbf{0.8917} & 0.8152 & 0.8171 & 0.8318 & \textbf{0.8471} & 0.7909 & 0.8027 & 0.8132 & 0.8120 & 0.8158 & 0.8199 \\
            & S+M-Feature (Proposed)                       & 0.8675 & 0.8717 & \textbf{0.8362} & \textbf{0.8335} & \textbf{0.8361} & 0.8389 & \textbf{0.8412} & \textbf{0.8415} & \textbf{0.8233} & \textbf{0.8228} & \textbf{0.8349} & \textbf{0.8342} \\
            \hline
            \multirow{4}{*} {Pre-training} & S$_{FR}$+M-Feature & 0.8175 & 0.8293 & 0.7390 & 0.7355 & 0.5930 & 0.6484 & 0.7705 & 0.7810 & 0.7484 & 0.7431 & 0.7441 & 0.7471 \\
            & S$_{KS+KD}$+M-Feature & 0.8140 & 0.8405 & 0.8314 & 0.8279 & 0.7500 & 0.7829 & 0.8073 & 0.8096 & 0.7792 & 0.7783 & 0.8033 & 0.8060 \\
            & S$^{Ko}_{Truth}$+M-Feature & 0.8534 & 0.8654 & 0.8258 & 0.8241 & 0.7768 & 0.7880 & 0.8248 & 0.8227 & 0.7837 & 0.7800 & 0.8102 & 0.8095 \\
            & S+M-Feature (Proposed)                       & \textbf{0.8675} & \textbf{0.8717} & \textbf{0.8362} & \textbf{0.8335} & \textbf{0.8361} & \textbf{0.8389} & \textbf{0.8412} & \textbf{0.8415} & \textbf{0.8233} & \textbf{0.8228} & \textbf{0.8349} & \textbf{0.8342} \\
            \hline
            \multirow{4}{*} {Loss} & L1 Loss baseline & 0.8094 & 0.8205 & 0.8162 & 0.8136 & 0.7266 & 0.7379 & 0.8113 & 0.7984 & 0.7918 & 0.7811 & 0.8011 & 0.7957 \\
            & SRCC Loss     & 0.8526 & 0.8632 & 0.8308 & 0.8296 & 0.7982 & 0.8120 & 0.8405 & \bf{0.8428} & 0.7931 & 0.7821 & 0.8192 & 0.8170 \\
			& PLCC Loss     & 0.8582 & 0.8563 & 0.8289 & 0.8275 & 0.8029 & 0.8226 & 0.8343 & 0.8423 & 0.8152 & 0.8104 & 0.8256 & 0.8260 \\
            & Mixed Loss (Proposed) & \textbf{0.8675} & \textbf{0.8717} & \textbf{0.8362} & \textbf{0.8335} & \textbf{0.8361} & \textbf{0.8389} & \textbf{0.8412} & 0.8415 & \textbf{0.8233} & \textbf{0.8228} & \textbf{0.8349} & \textbf{0.8342} \\
			\hline
			\multirow{4}{*} {Ensemble} & GRU & 0.8675 & 0.8717 & 0.8362 & 0.8335 & 0.8361 & 0.8389 & 0.8412 & \bf{0.8415} & 0.8233 & 0.8228 & 0.8349 & 0.8342 \\
            & Transformer      & 0.8527 & 0.8690 & 0.8382 & 0.8358 & 0.8303 & 0.8275 & 0.8355 & 0.8272 & 0.8193 & 0.8162 & 0.8318 & 0.8295 \\
            & GRU+Transformer & \textbf{0.8703} & \textbf{0.8760} & \textbf{0.8441} & \textbf{0.8394} & \textbf{0.8529} & \textbf{0.8467} & \textbf{0.8457} & 0.8386 & \textbf{0.8362} & \textbf{0.8291} & \textbf{0.8441} & \textbf{0.8388} \\
            & (Ensemble ratio)     & \multicolumn{2}{c|} {(0.67)} & \multicolumn{2}{c|} {(0.55)} & \multicolumn{2}{c|} {(0.65)} & \multicolumn{2}{c|} {(0.49)} & \multicolumn{2}{c|} {(0.53)} & \multicolumn{2}{c} {---} \\
            \bottomrule
	\end{tabular}}
	\label{Table:Ablation}
\end{table*}

\subsection{Ablation Study}\label{subsubsec:AblationStudy}
To verify the rationality of each module of the proposed method, we conduct a series of ablation studies from the following six aspects.

\textbf{Feature Ablation} We evaluate the effectiveness of different features fusion strategies. We begin with a {\bf{baseline}} which uses the pre-trained ResNet-50 on ImageNet as the frame-level feature extractor, which is equipped with a GRU to model the temporal information. We then replace the baseline frame-level feature extractor with a ResNet-50 trained using the proposed quality-aware pre-training scheme (\textbf{S-Feature}) or the motion features of SlowFast$_F$ (\textbf{M-Feature}). Note that the proposed method relies on both spatial and motion features (\textbf{S+M-Feature}), and all variants are trained with the mixed loss function as described in Section~\ref{Sec:LossFunction}. We report the results in the ``Feature'' section of Table~\ref{Table:Ablation}, where we have several interesting observations. First, S-Feature alone is able to introduce a $4.23\%$ improvement of weighted average SRCC over the baseline, indicating the effectiveness of the proposed quality-aware pre-training strategy. In contrast, M-Feature alone results in a significant performance drop on all databases, which suggests that spatial features play a vital role in perceiving the quality of videos. Combining S-Feature and M-Feature leads to the most promising results, demonstrating the complementarity between spatial and motion features for the VQA task. Typically, the motion features lead to a notable improvement (+$6.6\%$ SRCC) on LIVE-VQC, where motion-related distortions are prevailing~\cite{li2021unified, tu2021rapique}. We also replace GRU with average temporal pooling and then use SVR to learn a feature-quality mapping on S-Feature ({\bf{S-Feature+SVR}}). The unfavorable results highlight the importance of modeling temporal-memory effects in an appropriate way.

\textbf{Interaction Ablation} We also make two preliminary efforts to explore the spatio-temporal interaction effect. First, we integrate the spatial features generated from the slow pathway of SlowFast (dubbed as SlowFast$_S$) with our S+M-Feature (dubbed as S+M+SF$_S$-Feature). Profiting from the lateral connections between the two pathways in the SlowFast network, incorporating the features of the slow pathway is expected to capture the interaction between spatial and temporal distortions. Specifically, we up-sample the features of SlowFast$_S$ to the same temporal dimension of our S+M-Feature, followed by the channel-wise feature concatenation. Second, similar to~\cite{ying2021patch}, we aim for combining 2D and 3D pre-trained models to learn complementary features. Specifically, we use a pre-trained 3D ResNet-18 \cite{hara2017learning} to extract the spatio-temporal features  (dubbed as ST$_{3D}$-Feature). We down-sample the proposed S-Feature to the same temporal dimension of ST$_{3D}$-Feature, and then concatenate them (dubbed as S+ST$_{3D}$-Feature). As shown in the ``Interaction'' section of Table~\ref{Table:Ablation}, both practices are capable of introducing performance improvement in terms of weighted average SRCC and PLCC results over the S-Feature baseline, which further confirms the importance of combining spatial and temporal information for BVQA. However, neither of them outperforms the proposed S+M-Feature, which we believe is because the spatial features of the pre-trained SlowFast and the 3D ResNet-18 do not match the BVQA task well, resulting in negative transfer phenomena.

\textbf{Pre-training Ablation} We experiment with different quality-aware pre-training schemes. Specifically, similar to \cite{zhang2019blind}, we generate pseudo-labels for KADIS-700k \cite{lin2019kadid} using MS-SSIM \cite{wang2003multiscale}, and use them to pre-train a ResNet-50 in a regression manner (dubbed as \textbf{S$\mathbf{_{FR}}$}). We also follow the practice of \cite{wang2021rich} to train the frame-level feature extractor on KADIS-700k using a cross-entropy loss function and a pairwise hinge loss function, accounting for distortion types classification and degradation levels ranking. Note that we follow \cite{wang2021rich} to exclude distortion types 13 and 23 due to the license issue. After that, we fine-tune the pre-trained feature extractor on KADID-10k~\cite{lin2019kadid} with the above two loss functions and an L2 loss function for MOSs regression (dubbed as \textbf{S$_\mathbf{KS+KD}$}). As a comparison, we further use authentically distorted KonIQ-10k~\cite{hosu2020koniq} database with 10,073 human-annotated for quality-aware pre-training, dubbed as \textbf{S$\mathbf{^{Ko}_{Truth}}$}. The experimental results are summarized in the ``Pre-training'' section of Table~\ref{Table:Ablation}, from which we can observe that the proposed quality-aware pre-training strategy leads to the best performance, which we believe is due to the effective knowledge transfer from meaningful source domains with similar distortion scenarios (authentic distortions). Although \textbf{S$\mathbf{^{Ko}_{Truth}}$} is also trained with authentic distortions, the proposed strategy obtains more powerful feature representation due to incorporating broader realistic contents and more diverse distortions for quality-aware pre-training. We also notice that S$_{FR}$+M-Feature under-performs other competitors even with larger training samples, which suggests that the noisy proxy labels may mislead the representation learning, highlighting the importance of label precision.

\textbf{Loss Ablation} We evaluate the model trained with different loss functions, \textit{i.e.}, monotonicity-induced \textbf{SRCC Loss}, linearity-induced \textbf{PLCC Loss}, and the combination of them ({\bf{Mixed}}). Note that the L1 loss function is taken here as the baseline for comparison. All the variants are based on the S+M-Feature. In the ``Loss'' section of Table~\ref{Table:Ablation}, we find that both SRCC Loss and PLCC Loss alone are able to yield promising results, outperforming the L1 loss baseline by clear margins. By combining SRCC and PLCC loss functions, we obtain a 0.93\% additional gain in terms of the weighted SRCC. Notably, the Mixed Loss produces relatively significant improvements on CVD2014, LIVE-Qualcomm, and YouTube-UGC.

\begin{table*}[!htbp]
    \centering
	\caption{The CORAL~\cite{sun2016deep} distance between the source and target domains. A lower value of the CORAL distance indicates that the source and target domains are closer in the feature space. Note these values need to multiply $10^{-5}$. The last two columns show the median SRCC and PLCC results for overall performance}
	\setlength{\tabcolsep}{2.2mm}{
		\begin{tabular}{l|c|c|c|c|c|c|cc}
			\toprule
			\multirow{2}{*} {{\bf Database}} & \multirow{2}{*} {CVD2014} & \multirow{2}{*} {KoNViD-1k} & \multirow{2}{*} {LIVE-Qualcomm} & \multirow{2}{*} {LIVE-VQC} & \multirow{2}{*} {YouTube-UGC} & \multirow{2}{*} {Mixed VQA} & \multicolumn{2}{c} {W.A.} \\
            \cline{8-9}
            & & & & & & & SRCC & PLCC \\
			\hline
            ImageNet & 6.6480 & 2.6163 & 5.5639 & 2.6648 & 2.2154 & 1.8917 & 0.7726 & 0.7888 \\
            BID & 3.3920 & 1.7005 & 2.0990 & 1.4037 & 2.0297 & 1.4509 & 0.8126 & 0.8155 \\
            LIVEC & 2.6709 & 1.5320 & 1.7367 & 1.0034 & 1.6620 & 1.1324 & 0.8026 & 0.8132 \\
            KonIQ-10k & 3.1534 & 1.0601 & 1.7071 & 1.0335 & 1.3171 & 0.8833 & 0.8102 & 0.8095 \\
            SPAQ & 3.1084 & 1.7760 & 1.8764 & 1.5239 & 1.3138 & 1.2749 & 0.8183 & 0.8239\\
            Mixed IQA & 3.6190 & 1.2398 & 1.8628 & 1.2534 & 1.1482 & 0.9828 & 0.8349 & 0.8342 \\
            \bottomrule
	\end{tabular}}
	\label{Table:CoralDistance}
\end{table*}

\textbf{Ensemble Ablation} As an alternative to the \textbf{GRU}, we use a \textbf{Transformer}~\cite{vaswani2017attention} encoder to model the temporal information, where the number of layers, the dimension of the feed-forward network, the number of heads, and the dropout ratio are set to 2, 2048, 2, and 0.2, respectively. We also explore a simple ensemble trick to further boost the performance as:
\begin{equation}\label{Equation:EnsembleGRUTrans}
Q_e = \kappa Q_f^{G} + (1-\kappa) Q_f^{T}
\end{equation}
where $Q_f^{G}$ and $Q_f^{T}$ are the predicted scores with GRU and Transformer, respectively. The parameter $\kappa$ is the combination factor in the ensemble procedure. Concretely, we conduct experiments by varying $\kappa$ from 0 to 1, stepped by 0.01. $Q_e$ is the final ensemble quality score. As shown in the ``Ensemble'' section of Table~\ref{Table:Ablation}, GRU slightly outperforms Transformer, and an ensemble of them leads to a 0.92\% gain on the weighted average SRCC performance.

\textbf{Distance Ablation} It is interesting to explore whether we can rely on the CORAL~\cite{sun2016deep} distance to select source domains. To this end, we treat the mixed IQA databases as the source domain and compute the CORAL distances between it and VQA databases. The pairwise distance results are shown in Table~\ref{Table:CoralDistance}. We have several useful observations. First, we can rely on CORAL distance to select better source domains when the difference of distances is significant enough. For example, the distances between ImageNet and VQA databases are larger than all IQA domains, resulting in worse final performances. Second, the distances are not entirely monotonic to the overall performance on the target domains. This is a reasonable phenomenon because we measure covariate shifts between two domains with an assumption that the conditional distributions (\ie, from input to quality predictions) and label distributions (\ie, MOSs) of the source and target domains are exactly the same, only by which we can simplify measuring covariate shifts into measuring the marginal distribution shifts between the source and target domains. Unfortunately, this assumption is difficult to hold in practice, where the quality spaces of IQA and VQA tasks are not entirely identical (label distribution), and the conditional distributions (\ie, the mapping functions from images or videos to their quality spaces) are also different. Empirically, we achieve promising results on the target VQA tasks by combining multiple diverse IQA databases as the source domain.

\subsection{Computational Complexity}\label{subsec:ComputationAnalysis}
In practical applications, computational efficiency is desperately desired. We benchmark the computational complexity in this subsection. To make a fair comparison, all the methods are tested on the same machine, \ie, a Dell Precision 7920 Tower Workstation equipped with an Intel Xeon(R) Gold 5220R CPU$\times$2 @2.20Ghz$\times$96, 128G RAM, and NVIDIA Quadro RTX6000 24G GPU$\times$2. We use the implementations of the compared methods released by their authors. All methods are tested with MATLAB R2020a or Python 3.8.8, both under the Ubuntu 18.04.5 LTS operating system. We test our method using CPU and GPU, respectively. Meanwhile, we adopt both serial (SEL) and parallel (PAL) modes for our proposed two groups of features. These test methods are briefly recorded as Proposed\_CPU\_SEL, Proposed\_CPU\_PAL, Proposed\_GPU\_SEL, and Proposed\_GPU\_PAL.

\begin{figure}[htbp!]
    \centering
    \captionsetup{justification=centering}
    \subfloat[]{\includegraphics[width=0.98\columnwidth]{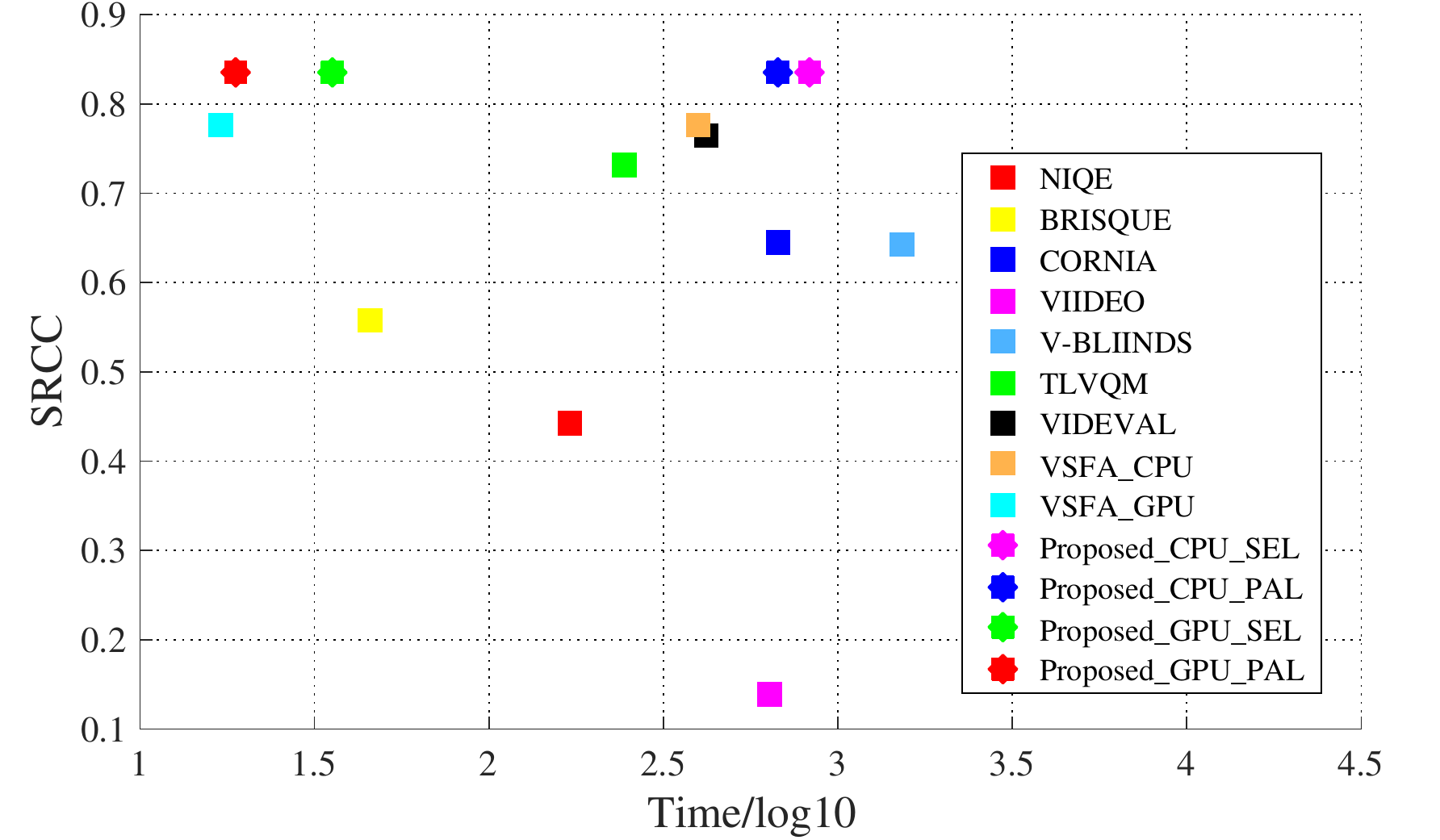}}\hskip.2em
    \subfloat[]{\includegraphics[width=1\columnwidth]{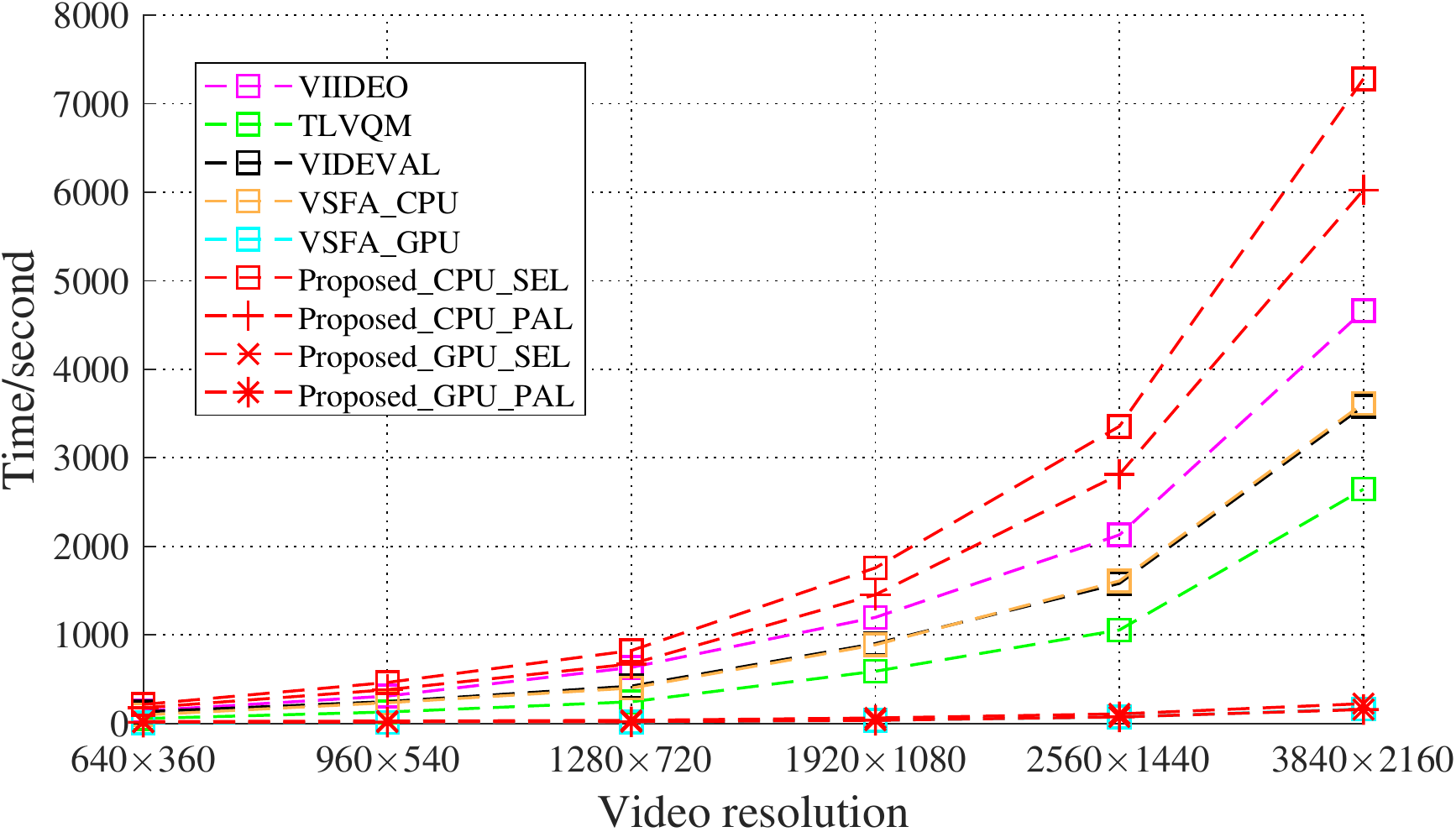}}
    \vspace{2pt}
    \caption{(a) The weighted average SRCC results (collected from Table~\ref{Table:SingleDatabasePerform_onCKLNY}) as a function of the running time in the logarithm space. (b) The running time as a function of the video resolutions.}\label{Figure:ComputationAnalysis}
\end{figure}

First, by fixing the resolution at 1280$\times$720, we plot the performance as a function of the runtime in Fig.~\ref{Figure:ComputationAnalysis}(a). Second, we evaluate the variation of runtime under different resolutions, \ie, 360p, 540p, 720p, 1080p, 1440p and 2160p, as displayed in Fig.~\ref{Figure:ComputationAnalysis}(b). All the results are derived from the average of ten repeated tests that aims to remove random bias. We select a video with a resolution of 1280$\times$720 and temporal length of 467 frames from CVD2014, and all videos with different resolutions are transformed from it. Overall, we have two important conclusions. First, Fig.~\ref{Figure:ComputationAnalysis}(a) shows that the proposed method (accelerated with GPU) achieves a favorable trade-off between effectiveness and efficiency. Second, as shown in Fig.~\ref{Figure:ComputationAnalysis}(b), the proposed method delivers a larger advantage in efficiency as the resolution increases. For example, with the resolution growing from 360p to 2160p, our algorithm (Proposed\_GPU\_PAL) can achieve an increase in speed from 3 times to 16 times compared against TLVQM. Although the proposed method has a higher computational complexity on CPU, it delivers better prediction accuracy result in stark contrast to other methods. In addition, it can directly benefit from significant acceleration by GPU. In the future, it would be a promising topic to compress the model for better efficiency.

\section{Conclusion}\label{Sec:Conclution}

We have proposed a DNN-based BVQA method for the in-the-wild scenario. We use model-based transfer learning methods to leverage knowledge from two types of source domains, corresponding to spatial appearance and temporal motion, respectively. Specifically, we conduct a quality-aware pre-training on multiple IQA databases to learn the frame-level feature extractor, which significantly enhances the feature representation without laborious efforts on video quality annotation. Similarly, we use a pre-trained DNN on action recognition to account for the motion perception of videos, which is complementary to the spatial features. We verify the promising performance of the proposed method through extensive experiments on six in-the-wild VQA databases. Besides, the merging of the differentiable PLCC and SRCC loss functions further boosts the performance.

As a limitation of the current model, it remains to be a challenging task to explore a more rational spatio-temporal interaction strategy. We believe the efficient joint optimization of the spatio-temporal representation is a promising direction. In addition, it is also important to incorporate the viewing conditions for making quality predictions~\cite{duanmu2021quantifying} of videos captured in varying environments. Another direction is developing effective continual learning methods~\cite{zhang2021continual,zhang2021task} for handling the BVQA where the data is streaming.

\bibliographystyle{IEEEtran}
\bibliography{ref}

\end{document}